\documentclass[12pt]{iopart}
\usepackage{iopams}  
\usepackage{amssymb}
\usepackage{bm}
\usepackage{epsfig}
\usepackage{wasysym}

\usepackage{color}

\begin{document}

\title{Universality classes of topological phase transitions with higher-order band crossing}

\author{Wei Chen$^{1}$ and Andreas P. Schnyder$^{2}$}

\address{$^{1}$Department of Physics, PUC-Rio, Rio de Janeiro 22451-900, Brazil}

\address{$^{2}$Max Planck Institute for Solid State Research, Stuttgart 70569, Germany}

\ead{wchen@puc-rio.br}
%\vspace{10pt}
%\begin{indented}
%\item[]August 2017
%\end{indented}

\begin{abstract}

In topological insulators and topological superconductors, the discrete jump of the topological invariant upon tuning a certain system parameter defines a topological phase transition. A unified framework is employed to address the quantum criticality of the topological phase transitions in one to three spatial dimensions, which simultaneously incorporates the symmetry classification, order of band crossing, $m$-fold rotational symmetry, correlation functions, critical exponents, scaling laws, and renormalization group approach. We first classify higher-order Dirac models according to the time-reversal, particle-hole, and chiral symmetries, and determine the even-oddness of the order of band crossing in each symmetry class. The even-oddness further constrains the rotational symmetry $m$ permitted in a symmetry class. Expressing the topological invariant in terms of a momentum space integration over a curvature function, the order of band crossing determines the critical exponent of the curvature function, as well as that of the Wannier state correlation function introduced through the Fourier transform of the curvature function. The conservation of topological invariant further yields a scaling law between critical exponents. In addition, a renormalization group approach based on deforming the curvature function is demonstrated for all dimensions and symmetry classes. Through clarification of how the critical quantities, including the jump of the topological invariant and critical exponents, depend on the nonspatial and the rotational symmetry, our work introduces the notion of universality class into the description of topological phase transitions.

\end{abstract}

%
% Uncomment for keywords
%\vspace{2pc}
%\noindent{\it Keywords}: XXXXXX, YYYYYYYY, ZZZZZZZZZ
%
% Uncomment for Submitted to journal title message
%\submitto{\JPA}
%
% Uncomment if a separate title page is required
%\maketitle
% 
% For two-column output uncomment the next line and choose [10pt] rather than [12pt] in the \documentclass declaration
%\ioptwocol
%

\section{Introduction}

A recently emerged issue in the research of topological insulators (TIs) and topological superconductors (TSCs) is the understanding of quantum criticality near topological phase transitions\cite{Continentino14,Continentino16,Kempkes16,Roy16,Griffith18,Chen17}. The topology of TIs and TSCs are characterized by the topological invariant defined from the Bloch wave function, which is calculated by different means according to the dimension and symmetry class of the system\cite{Schnyder08,Ryu10,Kitaev09,Chiu16}. The discrete jump of the topological invariant upon tuning a certain parameter $M$ at a critical point $M_{c}$ defines the topological phase transition. In practice, the low-energy effective theory of these materials are well described by a Dirac Hamiltonian that respects the symmetry of the system, and the tuning parameter $M$ represents the mass term, or equivalently the bulk gap of the energy spectrum, of the Dirac Hamiltonian. At the critical point $M_{c}$, the bulk gap closes, i.e., a topological phase transition necessarily involves a gap-closing. However, one should keep in mind that the converse is not true, namely gap-closing in the energy spectrum does not necessarily involve topology.

The first step towards understanding quantum criticality of such transitions is, of course, whether any asymptotic critical behavior manifests as $M\rightarrow M_{c}$. Given that the system does not necessarily possess a Landau order parameter, and the topological invariant jumps discretely at the critical point, the notion of asymptotic critical behavior that unambiguously defines the transition does not seem to be obvious at a glance. Recently, a series of work on scaling and critical exponents shed a light on this issue\cite{Chen16,Chen16_2,Chen17}. The observation in these works is that the topological invariant, being a global property of the Bloch state, is always calculated through momentum space integration over a certain function referred to as the curvature function. If $M$ is varied but the system remains the same topological phase, then the topological invariant remains unchanged, but the profile of the curvature function changes. Moreover, as $M\rightarrow M_{c}$, the curvature function gradually develops a divergence at the high symmetry point (HSP) where the gap-closing takes place. On the two sides of the critical point $M_{c}^{+}$ and $M_{c}^{-}$, the divergence is of opposite sign. This divergence and sign change of the curvature function serve as the asymptotic critical behavior that unambiguously identifies a topological phase transition, in constrast to any other quantum phase transition that does not involve topology.

Another major step towards understanding the topological phase transition is the identification of correlation functions, despite the system may not possess a local order parameter. Based on the theory of charge polarization\cite{KingSmith93,Resta94,Xiao10} and the theory of orbital magnetization\cite{Thonhauser05,Xiao05,Ceresoli06,Shi07,Souza08}, it is recognized that in certain symmetry classes, the Fourier transform of the curvature function yields a correlation function. In noninteracting systems, the correlation function is a measure of the overlap of Wannier functions that are a certain distance apart\cite{Chen17}. Because the curvature function diverges as $M\rightarrow M_{c}$, so does the correlation length $\xi$, consistent with the notion of scale invariance in the usual second-order phase transition. In addition, the power law divergence of $\xi$ with respect to $M$ defines the critical exponent $\nu$. It is further unveiled that the exponent $\nu$ is determined by the order of band crossing $n$ of the Dirac Hamiltonian at the critical point\cite{Chen17}. Depending on whether the bands at the critical point $M=M_{c}$ show linear, quadratic, or higher-order band crossing, the critical exponent and the jump of topological invariant take different values. These statements are drawn from investigating systems in a specific symmetry class, namely 2D class A in the periodic table of the topological materials, and is also found to manifest in 2D strongly interacting TIs solved by means of twisted boundary conditions\cite{Kourtis17}, and weakly interacting 1D and 2D TIs solved by means of single-particle Green's functions\cite{Chen18}.

On the other hand, previous investigation shows that an important mechanism to stabilize the aforementioned higher-order band crossing is the discrete $m$-fold rotational symmetry of the underlying lattice structure\cite{Fang12,Yang14}. Through investigating 2D and 3D multiple Weyl points
\cite{Xu11,Banerjee12,Lai15,Jian15,Huang15,DasSarma15,Huang16,Pyatkovskiy16,ChenFiete16,Ahn16}, 
the rotational eigenvalues are shown to be in one to one correspondence with the order of band crossing in the low-energy sector of these materials, which are described by higher-order Dirac or Weyl models\cite{Fang12,Yang14}. The inclusion of additional nonspatial symmetry further constrains the rotational eigenvalues and hence the band crossing\cite{Yang14}. However, shall this stabilization mechanism be generally applicable to TIs and TSCs in any symmetry class, it must also take into account the fact that the order of band crossing is already constrained by the nonspatial symmetries in the symmetry classification\cite{Schnyder08,Ryu10,Chiu16}.

In this article, we investigate the quantum criticality of topological phase transitions within a framework that simultaneously incorporates all the aforementioned ingredients, namely (1) symmetry classification, (2) order of band crossing, (3) $m$-fold rotational symmetry (for 2D and 3D), (4) correlation function, (4) critical exponents, and (5) the discrete jump of topological invariant. Our classification scheme is based on the following procedure. We first classify the higher-order Dirac models from 1D to 3D according to the time-reversal (TR), particle hole (PH), and chiral symmetry, and clarify the even-oddness of the order of the band crossing $n$ in each symmetry class. We reveal that the order of band crossing $n$ in certain classes also determines the jump of topological invariant $\Delta{\cal C}$ per gap-closing. When rotational symmetry is further imposed in 2D and 3D, the even-oddness of $n$ also constrains the rotational eigenvalues. As a result, the allowed rotational symmetry $m$ depends on the dimension, symmetry class, and the system being fermionic or bosonic. The notion of Wannier state correlation function is clarified for all the 15 topologically nonotrivial dimension-symmetry classes. Moreover, we verify that the critical exponent $\nu$ of the correlation length $\xi$, as well as that of the edge state decay length in the topologically nontrivial phase, is always determined by the order of band crossing $n$, but not necessarily $m$ or $\Delta{\cal C}$. On top of these, we will demonstrate that a previously proposed scaling law\cite{Chen17}, as well as a renormalization group approach based on the curvature function\cite{Chen16,Chen16_2}, remain valid for all the 15 cases. All these features together give rise to a coherent picture for the quantum criticality of topological phase transitions, especially how the critical quantities $\left\{\Delta{\cal C},n,\nu\right\}$ depend on the dimension and symmetry, which we refer to as the universality classes of TIs and TSCs.

The article is organized in the following manner. In Sec.~\ref{sec:generic_critical_behavior}, we give an overview of the generic critical behavior that is universal to all the 15 dimension-symmetry classes from 1D to 3D, including the divergence of the curvature function, critical exponents, scaling laws, and the renormalization group approach. In Secs.~\ref{sec:three_dimension} to \ref{sec:one_dimension}, we detail the interplay between TR, PH, chiral, as well as the $m$-fold rotational symmetry in 2D and 3D, especially how they influence the critical quantities $\left\{\Delta{\cal C},n,\nu\right\}$. The results are concisely summarized in Table \ref{tab:universality_class}. Because some classes in lower dimensions can be obtained from higher ones through a dimensional reduction\cite{Ryu10}, we start from 3D systems, and then reduce the dimension to 2D and 1D. Section \ref{sec:conclusions} summarizes the features obtained within this framework.

\begin{table}
\caption{Summary of the results in the 15 topologically nontrivial dimension-symmetry classes from 1D to 3D, which lists the topological invariant ${\cal C}$, jump of topological invariant across a topological phase transition per gap-closing $\Delta{\cal C}$, the order of band crossing $n$, the allowed $m$-fold rotational symmetry for fermionic $F=1$ and bosonic $F=0$ systems in 2D and 3D, and the corresponding Wannier state correlation function $\tilde{F}$.}
\footnotesize
\centering
\label{tab:universality_class}
\begin{tabular}{@{}lcccccc}
\br
 & ${\cal C}$ & $\Delta{\cal C}$ & $n$ & $m_{F=1}$ & $m_{F=0}$ & $\tilde{F}$ \\
\mr
1D class BDI & ${\mathbb Z}$ & 1 & $2{\mathbb Z}+1$ & & & $\tilde{F}_{1D}$ \\
1D class AIII & ${\mathbb Z}$ & 1 & ${\mathbb Z}$ & & & $\tilde{F}_{1D}$ \\
1D class DIII & ${\mathbb Z}_{2}$ & 1 & $2{\mathbb Z}+1$ & & & 
$\tilde{F}_{TR}$ \\
1D class D & ${\mathbb Z}_{2}$ & 1 & $2{\mathbb Z}+1$ & & & $\tilde{F}_{1D}$ \\
1D class CII & $2{\mathbb Z}$ & 2 & $2{\mathbb Z}+1$ & & & $\tilde{F}_{1D}^{\prime}$ \\
2D class A & ${\mathbb Z}$ & ${\mathbb Z}$ & ${\mathbb Z}$ & $2,3,4,6$ & $2,3,4,6$ & $\tilde{F}_{2D}$ \\
2D class C & $2{\mathbb Z}$ & $2{\mathbb Z}$ & $2{\mathbb Z}$ & $3$ & $2,3,4,6$ & $\tilde{F}_{2D}$ \\
2D class D & ${\mathbb Z}$ & $2{\mathbb Z}+1$ & $2{\mathbb Z}+1$ & $2,3,4,6$ & $3$ & $\tilde{F}_{2D}$ \\
2D class AII & ${\mathbb Z}_{2}$ & $1$ & $2{\mathbb Z}+1$ & $2,3,4,6$ & $2,3,4,6$ & $\tilde{F}_{TR}$ \\
2D class DIII & ${\mathbb Z}_{2}$ & 1 & $2{\mathbb Z}+1$ & $2,3,4,6$ & 3 & $\tilde{F}_{TR}$ \\
3D class AIII & ${\mathbb Z}$ & $2{\mathbb Z}+1$ & $2{\mathbb Z}+1$ & $2,3,4,6$ & $2,3,4,6$ & $\tilde{F}_{3D}$ \\
3D class DIII & ${\mathbb Z}$ & $2{\mathbb Z}+1$ & $2{\mathbb Z}+1$ & $2,3,4,6$ & 3 & $\tilde{F}_{3D}$ \\
3D class AII & ${\mathbb Z}_{2}$ & 1 & $2{\mathbb Z}+1$ & $2,3,4,6$ & 3 & $\tilde{F}_{TR}$ \\
3D class CII & ${\mathbb Z}_{2}$ & 1 & $2{\mathbb Z}+1$ & $2,3,4,6$ & 3 & $\tilde{F}_{3D}^{\prime}$ \\
3D class CI & $2{\mathbb Z}$ & $2{\mathbb Z}$ & $2{\mathbb Z}+1$ &  &  & $\tilde{F}_{3D}$ \\
\br
\end{tabular}\\
\end{table}
\normalsize

\section{Generic critical behavior of higher-order Dirac models \label{sec:generic_critical_behavior}}

\subsection{Curvature function and correlation function \label{sec:curvature_fn_correlation_fn}}

We discuss the TIs and TSCs whose low-energy effective theory is a Dirac model satisfying the following criteria:

\begin{flushleft}

%The $\Gamma^{ab}$ matrices, which may account for effects such as spin-orbit coupling, are not included. 

(1) The low-energy effective Dirac Hamiltonian $H({\bf k})={\bf d}({\bf k})\cdot{\boldsymbol\Gamma}$ is constructed out of $\Gamma^{a}$ matrices that satisfies the Clifford algebra $\left\{\Gamma^{a},\Gamma^{b}\right\}=2\delta_{ab}$. In addition, the low-energy dispersion takes the form
\begin{eqnarray}
E_{\pm}({\bf k})=\pm d=\pm\left(k^{2n}+M^{2}\right)^{1/2}\;,
\label{general_Dirac_dispersion}
\end{eqnarray}
where the integer $n$ signifies the order of band crossing at the topological phase transition $M_{c}=0$. The mass term $M$ is assumed to have no momentum dependence. 

(2) The model is classfied according to the TR, PH and chiral symmetries 
\begin{eqnarray}
&&TH({\bf k})T^{-1}=H(-{\bf k})\;,
\nonumber \\
&&CH({\bf k})C^{-1}=-H(-{\bf k})\;,
\nonumber \\
&&SH({\bf k})S^{-1}=-H({\bf k})\;.
\label{General_symmetries_TR_PH_CH}
\end{eqnarray}
In each spatial dimension from 1D to 3D, there are 5 symmetry classes that are topologically nontrivial. We will discuss all the 15 nontrivial dimension-symmetry classes case by case in the following sections.

(3) In 2D and 3D, we consider $m$-fold rotation symmetric Dirac Hamiltonians and choose an eigenvalue basis in which both the Dirac Hamiltonian and the rotation operator are diagonal. In this basis, the operator of the $m$-fold rotation symmetry can be written as
\begin{eqnarray}
C_{m}={\rm diag}(\alpha_{p},\alpha_{q}...)\;,\;\;\;\alpha_{p}=e^{i\frac{2\pi}{m}\left(p+\frac{F}{2}\right)}\;,
\end{eqnarray}
where $p=0,1,2...m-1$ are the rotational eigenvalues. The spin statistical factor $F$ indicates whether the basis is fermionic $F=1$ or bosonic $F=0$. We assume that in each symmetry class, the basis can be either fermionic or bosonic. The rotational symmetry requires that 
\begin{eqnarray}
C_{m}H({\bf k})C_{m}^{-1}=H(R_{m}{\bf k})\;,
\label{general_rotational_symmetry}
\end{eqnarray}
where $R_{m}$ rotates the momentum ${\bf k}$ about the ${\bf k}_{z}$ axis by $2\pi/m$. In 1D this rotational symmetry is absent.

\end{flushleft}

As will be demonstrated case by case in the following sections, the topological invariant ${\cal C}$ (or a related one that judges the topological phase transition) in all the 15 dimension-symmetry classes can be cast into the form of a $D$-dimensional integration over the Brillouin zone (BZ)
\begin{eqnarray}
{\cal C} = {\cal C}(M) =\int_{\rm BZ} \frac{d^{D}{\bf k}}{(2\pi)^{D}}\;F({\bf k},M)\;,
\label{general_topological_invariant_curvature}
\end{eqnarray}
where $F({\bf k},M)$ is referred to as the curvature function, 
the precise form of which depends on the dimension and symmetry class. The points in the BZ satisfying ${\bf k}_{0}=-{\bf k}_{0}$ (up to a reciprocal lattice vector) are referred to as the high symmetry points (HSPs). The curvature function is generally an even function $F({\bf k}_{0}+\delta{\bf k},M)=F({\bf k}_{0}-\delta{\bf k},M)$ around the HSP due to certain symmetry, such as inversion or TR symmetry. While the topological invariant ${\cal C}$ remains constant within a topological phase in the parameter space of $M$, the profile of $F({\bf k},M)$ varies with changing $M$, which is key to our analysis of critical behavior.

Another key ingredient in our analysis is the Wannier states $|{\bf R} n\rangle$ defined from the Bloch state by 
\begin{eqnarray}
|u_{n{\bf k}}\rangle=\sum_{{\bf R}}e^{-i {\bf k}\cdot({\hat{\bf r}}-{\bf R})}|{\bf R} n\rangle\;,\;\;\;
|{\bf R} n\rangle=\frac{1}{N}\sum_{\bf k}e^{i {\bf k}\cdot({\hat{\bf r}}-{\bf R})}|u_{n{\bf k}}\rangle\;,
\label{Wannier_basis}
\end{eqnarray}
where $ N $ denotes the number of lattice sites, and ${\hat{\bf r}}$ is the position operator. We propose a correlation function derived through the Fourier transform of the curvature function, 
\begin{eqnarray}
\lambda_{\bf R}=\int \frac{d^{D}{\bf k}}{(2\pi)^{D}}\;e^{i{\bf k}\cdot{\bf R}}\;F({\bf k},M)\;.
\label{general_correlation_function}
\end{eqnarray}
In each of the 15 dimension-symmetry classes, we will demonstrate that the correlation function always takes the form of measuring the overlap of Wannier functions.

\subsection{Generic critical behavior and universality classes \label{sec:scaling_hypothesis}}

Through investigating the low-energy continuous models of all the 15 nontrivial dimension-symmetry classes from 1D to 3D, we found that the critical behavior of the curvature functions fall into two different scenarios:

\emph{Peak-divergence scenario.--} For linear band crossing $n=1$, the curvature function $F({\bf k},M)$ peaks at the gap-closing HSP. The peak gradually narrows and the height increases as the system approaches the critical point $M\rightarrow M_{c}$, and eventually the peak diverges and flips sign across the transition. We call this critical behavior the peak-divergence scenario. In this scenario, the peak can be well fitted by an Ornstein-Zernike form 
\begin{eqnarray}
F({\bf k}_{0}+\delta{\bf k},M)=\frac{F({\bf k}_{0},M)}{1+\xi^{2}\delta k^{2}}\;.
\end{eqnarray}
The critical behavior described above is summarized as 
\begin{eqnarray}
&&\lim_{M\rightarrow M_{c}^{+}}F({\bf k}_{0},M)=-\lim_{M\rightarrow M_{c}^{-1}}F({\bf k}_{0},M)=\pm\infty\;,\;\;\;
\nonumber \\
&&\lim_{M\rightarrow M_{c}}\xi=\infty\;.
\end{eqnarray}
Denoting the critical exponents of $F({\bf k}_{0},M)$ and $\xi$ as 
\begin{eqnarray}
|F({\bf k}_{0},M)|\propto|M-M_{c}|^{-\gamma}\;,\;\;\;
\xi\propto|M-M_{c}|^{-\nu}\;,
\end{eqnarray}
the conservation of the topological invariant ${\cal C}_{div}$ under the diverging peak 
\begin{eqnarray}
{\cal C}_{\rm div}=F({\bf k}_{0},M)\left(\prod_{i=1}^{D}\int_{-\xi^{-1}}^{\xi^{-1}}\frac{dk_{i}}{1+\xi^{2}k_{i}^{2}}\right)=\frac{F({\bf k}_{0},M)}{\xi^{D}}\times{\cal O}(1)={\rm const.}\;,\;\;\;
\label{Cdiv_peak_divergent}
\end{eqnarray}
yields a scaling law 
\begin{eqnarray}
\gamma=D\nu\;,
\label{scaling_law_peak_divergence}
\end{eqnarray}
which is to be satisfied by any linear, isotropic Dirac model in any dimension. Because of this Ornstein-Zernike form, the Wannier state correlation function as a Fourier transform of the curvature function, as in Eq.~(\ref{general_correlation_function}), decays with correlation length $\xi$.

\begin{figure}[ht]
\begin{center}
\includegraphics[clip=true,width=0.75\columnwidth]{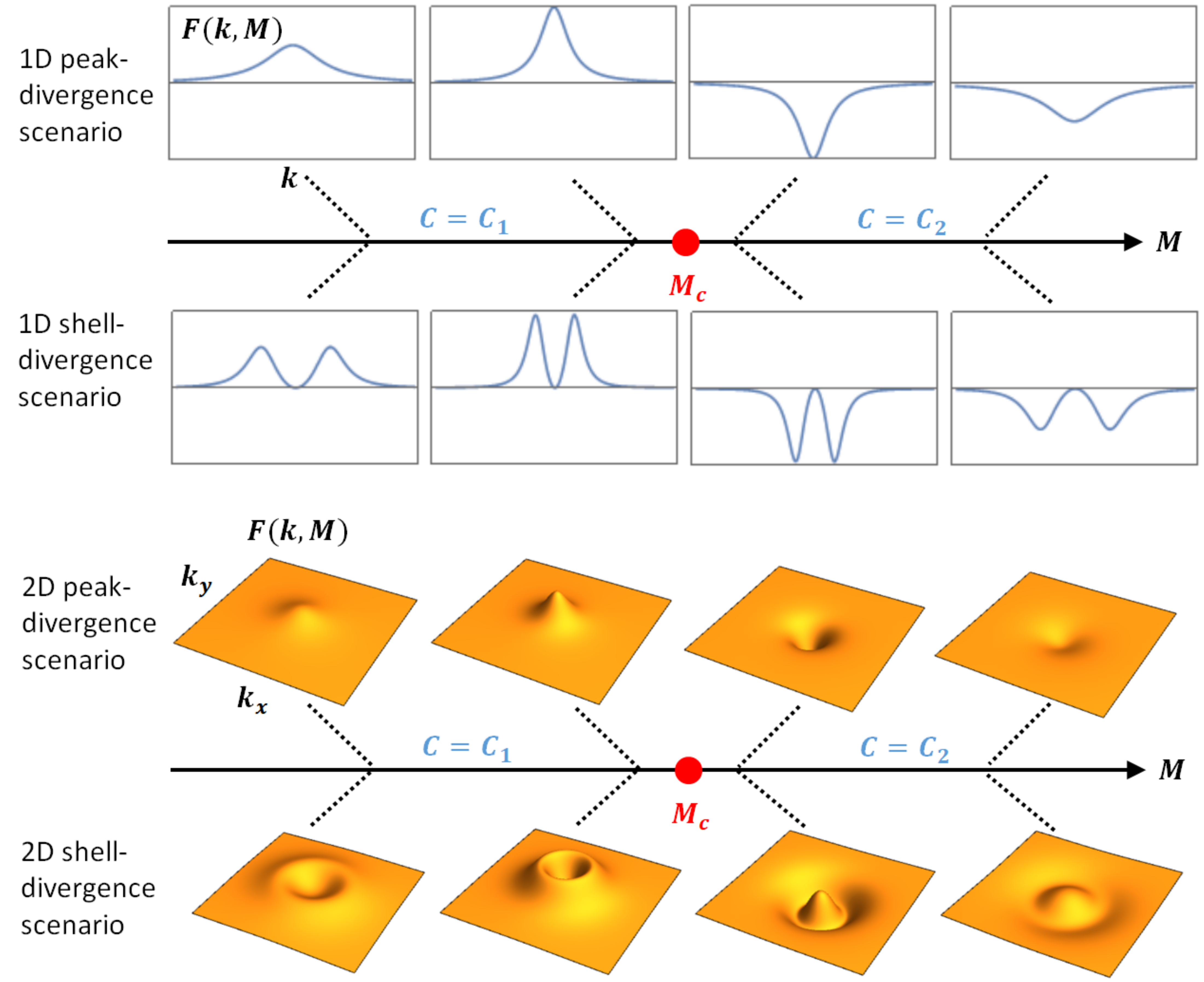}
\caption{ Schematics of the evolution of the curvature function $F({\bf k},M)$ in 1D (top) and 2D (bottom) near the topological phase transition $M_{c}$. The curvature function near the HSP is shown schematically for both close to and far away from the critical point $M_{c}$. In the peak-divergence scenario, the $F({\bf k},M)$ displays a Lorentzian peak at the HSP that becomes sharper as $M\rightarrow M_{c}$, and changes sign across $M_{c}$. In contrast, in the shell-divergence scenario the $F({\bf k},M)$ has an extremum at momentum away from the HSP, whose radius reduces and magnitude increases as $M\rightarrow M_{c}$. The extremum changes sign across $M_{c}$. Within the same topological phase, despite the profile of $F({\bf k},M)$ changes with $M$, the volume underneath $F({\bf k},M)$ is the topological invariant and hence remains unchanged, leading to the scaling laws. } 
\label{fig:peak_divergence_ring_divergence}
\end{center}
\end{figure}

\emph{Shell-divergence scenario.--} For any high order band crossing $n>1$, the extremum of the curvature function $F_{max}({\bf k}_{max},M)$ forms a $D-1$ dimensional shell surrounding the HSP, with a radius denoted by $k_{max}$ and a thickness denoted by $k_{wid}$. The critical behavior is that the shell gradually reduces it radius as $M\rightarrow M_{c}$, and $F_{max}({\bf k}_{max},M)$ gradually diverges and flips sigh across the transition. In other words, 
\begin{eqnarray}
&&\lim_{M\rightarrow M_{c}^{+}}F_{max}({\bf k}_{max},M)=-\lim_{M\rightarrow M_{c}^{-1}}F_{max}({\bf k}_{max},M)=\pm\infty\;,\;\;\;
\nonumber \\
&&\lim_{M\rightarrow M_{c}}k_{max}=0\;,\;\;\;\lim_{M\rightarrow M_{c}}k_{wid}=0\;.
\end{eqnarray}
For the Dirac models in Sec.~\ref{sec:curvature_fn_correlation_fn}, the $k_{max}$ and $k_{wid}$ have the same critical exponent
\begin{eqnarray}
|F({\bf k}_{0},M)|\propto|M-M_{c}|^{-\gamma}\;,\;\;\;
k_{max}^{-1}\propto k_{wid}^{-1}\propto\xi\propto|M-M_{c}|^{-\nu}\;.
\end{eqnarray}
The conservation of the portion of the topological invariant ${\cal C}_{div}$ in the diverging shell 
\begin{eqnarray}
{\cal C}_{div}=\Omega\int_{k_{max}-k_{wid}/2}^{k_{max}+k_{wid}/2}dk\,k^{D-1}F({\bf k},M)\approx \Omega k_{wid}k_{max}^{D-1}F_{max}={\rm const}.
\end{eqnarray}
where $\Omega$ takes into account the angular integration, yields the same scaling law as Eq.~(\ref{scaling_law_peak_divergence}). The two scenarios are depicted pictorially for 1D and 2D in Fig.~\ref{fig:peak_divergence_ring_divergence}.

To be more specific, find that the low-energy continuous model for most of the symmetry classes has the topological invariant in Eq.~(\ref{general_topological_invariant_curvature}) in the following scaling form
\begin{eqnarray}
{\cal C}=\Omega\int_{0}^{\infty}dk\,k^{D-1}\frac{Mk^{D(n-1)}}{\left(M^{2}+k^{2n}\right)^{(D+1)/2}}\;.
\end{eqnarray}
Expanding the curvature function in the integrand for small $k$ yields 
\begin{eqnarray}
F(k,M)=\frac{Mk^{D(n-1)}}{\left(M^{2}+k^{2n}\right)^{(D+1)/2}}\approx
\frac{{\rm Sgn}(M)}{|M|^{D}}\frac{k^{D(n-1)}}{\left(1+\frac{D+1}{2}\frac{k^{2n}}{M^{2}}\right)}\;.
\label{general_FkM_scaling_form}
\end{eqnarray}
For the peak-divergence scenario at $n=1$, the Ornstein-Zernike form is evident, which renders $\gamma=D$ and $\nu=1$, satisfying Eq.~(\ref{scaling_law_peak_divergence}). For the shell-divergence scenario at $n>1$, solving for the $k_{max}$ from $\partial_{k}F(k,M)=0$ yields $k_{max}\propto|M|^{1/n}$, and subsequently we obtain $F_{max}(k_{max},M)\propto |M|^{-D/n}$. Consequently, the critical exponents are $\gamma=D/n$ and $\nu=1/n$, in agreement with the scaling law in Eq.~(\ref{scaling_law_peak_divergence}). This analysis indicates that the critical exponent $\nu$ for the length scale $\xi$ are basically determined by the order of band crossing $n$, as can also be inferred by counting the dimension in the dispersion in Eq.~(\ref{general_Dirac_dispersion}).

%{\cblue (1) The 2D class AII seems to be the only class that does not satisfy this scaling form. Need to check. }

The edge state decay length in the topologically nontrivial phase has the same critical exponent as the correlation length. This can be seen by considering the model to be defined in the half-space $x>0$, and considering the edge state with zero transverse momentum, i.e., $k_{y}=k_{z}=0$ in 3D and $k_{y}=0$ in 2D. The problem can be calculated by projecting the higher-order Dirac Hamiltonian into real space and solve for the zero-energy edge state\cite{Chen17}
\begin{eqnarray}
\left({\bf d}\cdot{\boldsymbol\Gamma}\right)\psi=\left[k_{x}^{n}\Gamma^{x}+M\Gamma^{M}\right]\psi
=\left[(-i)^{n}\partial_{x}^{n}\Gamma^{x}+M\Gamma^{M}\right]\psi=0\;,
\end{eqnarray} 
where $\Gamma^{x}$ and $\Gamma^{M}$ represent the $\Gamma$-matrices that multiply $k_{x}^{n}$ and $M$, respectively, whose precise form depends on the symmetry class and dimension. Multiplying the equation by $\Gamma^{x}$ and using the ansatz $\psi=\chi_{\eta}\phi(x)$, with $\chi_{\eta}$ an eigenstate of $\Gamma^{x}\Gamma^{M}\chi_{\eta}=\eta\chi_{\eta}$, we obtain 
\begin{eqnarray}
\xi^{n}=\frac{i^{n-2}}{\eta M}\;.
\end{eqnarray}
The eigenvalue $\eta$ is chosen such that one of the roots $\xi_{\rm loc}$ is real and identifiable with the edge state decay length, which obviously also has critical exponent $\nu=1/n$. Thus, the correlation length and the edge state decay length are essentially the same length scale, although we should emphasize that the edge state only exists in the topologically nontrivial phase, whereas the Wannier state correlation function is well-defined in either the trivial or nontrivial phase.

\subsection{Curvature renormalization group approach \label{sec:curvature_RG}}

\begin{figure}
\centering
\includegraphics[width=0.7\linewidth]{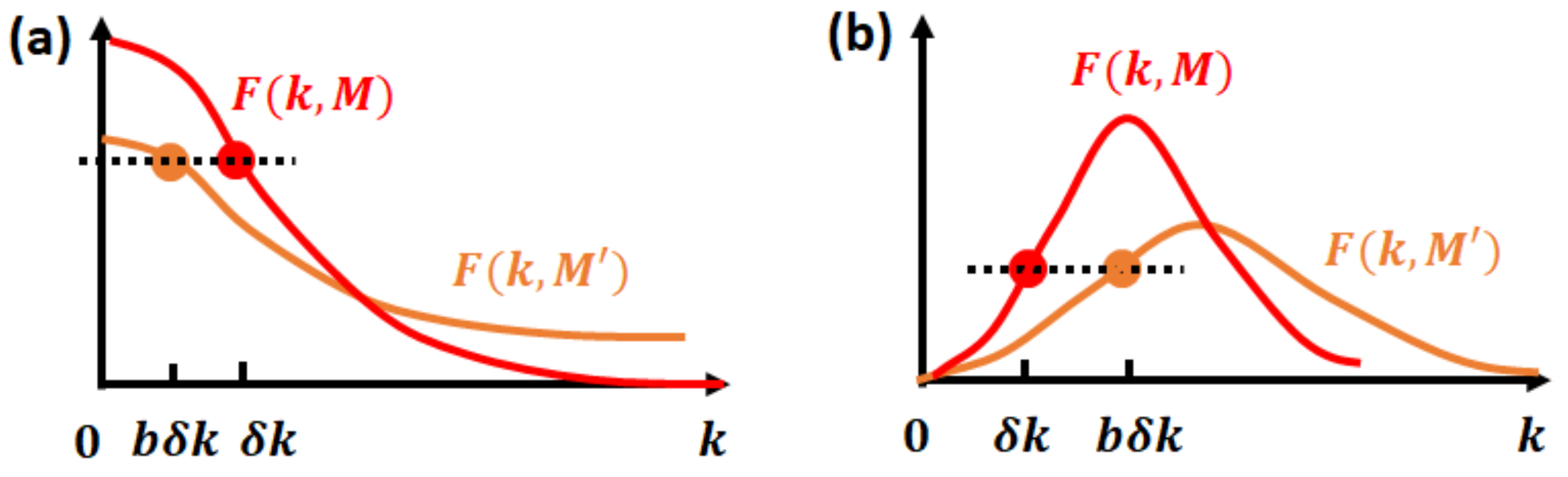}
\caption{Schematics of the CRG approach for (a) the peak-divergence scenario in which one chooses $b<1$, and for (b) the ring-divergence scenario in which one chooses $b>1$.
}
\label{fig:CRG_schematics}
\end{figure}

The curvature renormalization group (CRG) approach has been proposed to judge topological phase transitions in a variety of systems\cite{Chen16,Chen16_2,Chen17,Kourtis17,Chen18}. Here we demonstrate that the method applies to all the 15 dimension-symmetry classes at any order of band crossing $n$ and rotational symmetry $m$. The approach is essentially an iterative method to search for the trajectory (RG flow) in the parameter space of $M$ along which the maximum of the curvature function is reduced. In this way, the topological phase transitions, at which the curvature function diverges, can be identified. The scaling equation demands that, at a given $M$, we find the new $M'$ that satisfies
\begin{eqnarray}
F({\bf k}_{0}+\delta{\bf k},M)=F({\bf k}_{0}+b\delta{\bf k},M')\;.
\label{general_scaling_eq}
\end{eqnarray}
The mapping $M\rightarrow M'$ yields the RG flow that identifies the topological phase transitions. Here $\delta{\bf k}$ is a small momentum displacement away from the HSP, and one chooses $0<b<1$ for the peak-divergence scenario $n=1$, whereas $b>1$ for the shell-divergence scenario $n>1$. The divergence of the curvature function is gradually reduced under the scaling procedure, as shown schematically in Fig.~\ref{fig:CRG_schematics}. We remark that a similar RG approach has also been proposed to search for the transition based on the reduced density matrix obtained from bipartition, since it shows a similar critical behavior\cite{vanNieuwenburg18}.

Since the curvature function in many symmetry classes takes the generic scaling form of Eq.~(\ref{general_FkM_scaling_form}), we examine here explicitly the CRG approach applied to this scaling form. As the approach concerns only a small $k$ region near the HSP which is set to be at the origin in our continuous models ${\bf k}_{0}={\bf 0}$, we expand the scaling form in Eq.~(\ref{general_FkM_scaling_form}) by 
\begin{eqnarray}
F(\delta k,M)\approx
\frac{{\rm Sgn}(M)}{|M|^{D}}\delta k^{D(n-1)}\left(1-\frac{D+1}{2}\frac{\delta k^{2n}}{M^{2}}\right)\;.
\label{general_FkM_scaling_form_expansion}
\end{eqnarray}

For the peak-divergence scenario $n=1$, putting Eq.~(\ref{general_FkM_scaling_form_expansion}) into the leading order expansion of the scaling equation, Eq.~(\ref{general_scaling_eq}), yields the RG equation 
\begin{eqnarray}
\frac{dM}{d\ell}\equiv\frac{M'-M}{\delta k^{2}}=\frac{1-b^{2}}{2}\frac{\partial_{k}^{2}F(k,M)|_{k=0}}{\partial_{M}F(k,M)|_{k=0}}
=\frac{(1-b^{2})(D+1)}{2D}\frac{1}{M}\;.
\end{eqnarray}
The corresponding RG flow flows away from the critical point $M_{c}=0$, with a flow rate that diverges as $M\rightarrow M_{c}$.

For the shell-divergence scenario at $n>1$, we approximate Eq.~(\ref{general_FkM_scaling_form_expansion}) by 
\begin{eqnarray}
&&F(\delta k,M)\approx
\frac{{\rm Sgn}(M)}{|M|^{D}}\delta k^{D(n-1)}\;,
\nonumber \\
&&F(b\delta k,M')=F(\delta k+\Delta k,M+\delta M)\approx
\frac{{\rm Sgn}(M)}{|M+\delta M|^{D}}(\delta k+\Delta k)^{D(n-1)}\;.
\label{general_FkM_scaling_form_expansion_further}
\end{eqnarray}
Treating $\delta M\ll M$ and $\Delta k=(b-1)\delta k\ll\delta k$ as small quantities and expanding to the leading order, and then putting Eq.~(\ref{general_FkM_scaling_form_expansion_further}) into the Eq.~(\ref{general_scaling_eq}) yields the leading order RG equation 
\begin{eqnarray}
\frac{dM}{d\ell}\equiv\frac{\delta M}{(\Delta k/\delta k)}=(n-1)M\;.
\end{eqnarray}
The equation corresponds to an RG flow that flows away from $M_{c}=0$, with a flow rate that vanishes at $M=M_{c}$. In other words, the topological phase transition manifests as an unstable fixed point at $n>1$.

\section{Topological phase transitions in three dimensions \label{sec:three_dimension}}

In this section, we study topological phase transitions in three dimensions. We consider all the five symmetry classes that have nontrivial topology. The Wannier state correlation functions of class AIII, DIII, CII, and CI are constructed in a similar manner from the winding numbers, whereas in class AII it is constructed differently from the ${\mathbb Z}_{2}$ invariant.

\subsection{3D class AIII \label{sec:3D_class_AIII}}

For 3D class AIII that has $S^{2}=1$, we choose the $\Gamma$-matrices\cite{Ryu10} 
\begin{eqnarray}
&&\Gamma^{a}=\left(\alpha_{x},\alpha_{y},\alpha_{z},\beta,-i\beta\gamma^{5}\right)\;,
\nonumber \\
&&\alpha_{i}=\left(
\begin{array}{cc}
0 & \sigma_{i} \\
\sigma_{i} & 0
\end{array}
\right)\;,\;\;\;
\beta=\left(
\begin{array}{cc}
1 & 0 \\
0 & -1
\end{array}
\right)\;,\;\;\;
\gamma^{5}=\left(
\begin{array}{cc}
0 & 1 \\
1 & 0
\end{array}
\right)\;.
\label{3D_class_AIII_gamma_matrices}
\end{eqnarray}
The chiral operator is $S=\beta$, which together with the chiral symmetry in Eq.~(\ref{General_symmetries_TR_PH_CH}) requires that $d_{4}({\bf k})=0$, whereas there is no constraint on $d_{i}({\bf k})$ being even or odd in ${\bf k}$ for $i=1,2,3,5$. We follow the convention to choose $d_{5}=M$ as the mass term. The Hamiltonian then takes the form
\begin{eqnarray}
H({\bf k})=\sum_{i=1,2,3,5}d_{i}\Gamma^{i}=
\left(
\begin{array}{cccc}
 & & g^{\prime} & f \\
 & & f^{\ast} & -g^{\prime\ast} \\
g^{\prime\ast} & f & & \\
f^{\ast} & -g^{\prime} & & 
\end{array}
\right)\;,
\label{3D_class_AIII_Hamiltonian_general}
\end{eqnarray}
where we have denoted
\begin{eqnarray}
&&f=d_{1}-id_{2}=k_{+}^{n_{+}}k_{-}^{n_{-}}\;,
\nonumber \\
&&g^{\prime}=d_{3}-iM=g-iM=k_{z}^{n}-iM\;,
\label{3D_class_AIII_f_g}
\end{eqnarray}
with $n=n_{+}+n_{-}$, such that the dispersion satisfies
\begin{eqnarray}
E_{\pm}({\bf k})=\pm d=\pm\sqrt{|f|^{2}+|g|^{2}+M^{2}}
=\pm\sqrt{k^{2n}+M^{2}}\;,
\label{3D_class_AIII_Ek}
\end{eqnarray}
with $n\in{\mathbb Z}$.

Using the rotational symmetry operator 
\begin{eqnarray}
C_{m}={\rm diag}(\alpha_{p},\alpha_{q},\alpha_{r},\alpha_{s})\;,\;\;\;\alpha_{p}=e^{i\frac{2\pi}{m}\left(p+\frac{F}{2}\right)}\;,
\label{3D_class_AIII_Cm}
\end{eqnarray}
the chiral symmetry $[C_{m},S]=0$ does not give any constraint on $C_{m}$.
The rotational symmetry in Eq.~(\ref{general_rotational_symmetry}) demands 
\begin{eqnarray}
\alpha_{p}\alpha_{s}^{\ast}f(k_{\pm},k_{z})=\alpha_{r}\alpha_{q}^{\ast}f(k_{\pm},k_{z})=f(k_{\pm}e^{\pm i\frac{2\pi}{m}},k_{z})\;,
\nonumber \\
\alpha_{p}\alpha_{r}^{\ast}g'(k_{\pm},k_{z})=\alpha_{s}\alpha_{q}^{\ast}g'(k_{\pm},k_{z})=g'(k_{\pm}e^{\pm i\frac{2\pi}{m}},k_{z})\;.
\label{3D_class_AIII_f_g_constraint}
\end{eqnarray}
Because $g'=g-iM$ contains the mass term that must be invariant under the rotational transformation, it is required that $\alpha_{p}\alpha_{r}^{\ast}=1$ and $\alpha_{s}\alpha_{q}^{\ast}=1$, and therefore $r=p$ and $q=s$. The constraint on $g'$ then gives $g(k_{\pm},k_{z})=g(k_{\pm}e^{\pm i\frac{2\pi}{m}},k_{z})$, which is consistent with our parametrization in Eq.~(\ref{3D_class_AIII_f_g}). The parametrization of $f$ in Eq.~(\ref{3D_class_AIII_f_g}), together with $\alpha_{p}\alpha_{s}^{\ast}f(k_{\pm},k_{z})=\alpha_{p}\alpha_{q}^{\ast}f(k_{\pm},k_{z})=f(k_{\pm}e^{\pm i\frac{2\pi}{m}},k_{z})$ gives 
\begin{eqnarray}
n_{+}-n_{-}=xm+(p-q)\in{\mathbb Z}\;,
\label{3D_class_AIII_npmnm_pq}
\end{eqnarray}
which is the relation between the parametrization and the rotational symmetry eigenvalues, and we have used $n_{+}-n_{-}\in{\mathbb Z}$ due to $n=n_{+}+n_{-}\in{\mathbb Z}$. The $x$ in Eq.~(\ref{3D_class_AIII_npmnm_pq}) and throughout the article is an arbitrary integer. Evidently, Eq.~(\ref{3D_class_AIII_npmnm_pq}) can be satisfied by any of the $m=2,3,4,6$ for either the bosonic or fermionic basis.

We define the $q$-matrix by (not to be confused with one of the rotational eigenvalues)
\begin{eqnarray}
q({\bf k})=-\frac{1}{d}\left(d_{i}\sigma_{i}-iM\right)
=-\frac{1}{d}\left(
\begin{array}{cc}
g-iM & f \\
f^{\ast} & -g-iM
\end{array}
\right)
\;.
\end{eqnarray}
Writting the parametrization in Eq.~(\ref{3D_class_AIII_f_g}) into cylindrical coordinate $(k_{\perp},\phi,k_{z})$ by $f=k_{\perp}^{n}e^{i(n_{+}-n_{-})\phi}$ and $g=k_{z}^{n}$, and using the derivatives
\begin{eqnarray}
\frac{\partial}{\partial k_{x}}=\cos\phi\frac{\partial}{\partial k_{\perp}}-\frac{\sin\phi}{k_{\perp}}\frac{\partial}{\partial\phi}\;,\;\;\;
\frac{\partial}{\partial k_{y}}=\sin\phi\frac{\partial}{\partial k_{\perp}}+\frac{\cos\phi}{k_{\perp}}\frac{\partial}{\partial\phi}\;,
\end{eqnarray}
yields the following combination in the integration of topological invariant\cite{Ryu10}
\begin{eqnarray}
\epsilon^{\mu\nu\rho}{\rm Tr}\left[q^{\dag}\partial_{\mu}qq^{\dag}\partial_{\nu}qq^{\dag}\partial_{\rho}q\right]
=-M\frac{12k_{\perp}^{2(n-1)}k_{z}^{n-1}(n_{+}-n_{-})n^{2}}
{\left(k^{2n}+M^{2}\right)^{2}}\;.
\label{3D_class_AIII_Trqdq_cylindrical}
\end{eqnarray}
Converting the integrand into spherical coordinates $(k,\theta,\phi)$ yields 
\begin{eqnarray}
&&\epsilon^{\mu\nu\rho}{\rm Tr}\left[q^{\dag}\partial_{\mu}qq^{\dag}\partial_{\nu}qq^{\dag}\partial_{\rho}q\right]
\nonumber \\
&&=-M\frac{12(k\sin\theta)^{2(n-1)}(k\cos\theta)^{n-1}(n_{+}-n_{-})n^{2}}
{\left(k^{2n}+M^{2}\right)^{2}}\;.
\label{3D_class_AIII_Trqdq_spherical}
\end{eqnarray}
The topological invariant is given by the 3D winding number obtained from the integration of the above quantity 
\begin{eqnarray}
{\cal C}&=&\frac{\pi}{3}\int\frac{d^{3}{\bf k}}{(2\pi)^{3}}\epsilon^{\mu\nu\rho}{\rm Tr}\left[q^{\dag}\partial_{\mu}qq^{\dag}\partial_{\nu}qq^{\dag}\partial_{\rho}q\right]
\nonumber \\
&=&\left\{
\begin{array}{ll}
-\frac{(n_{+}-n_{-})}{2}{\rm Sgn}(M)\left[2nB(\frac{n}{2},n)\right] & {\rm if}\;n\in 2{\mathbb Z}+1\;, \\
0 & {\rm if}\;n\in 2{\mathbb Z}\;,
\end{array}
\right.
\label{3D_class_AIII_topo_inv}
\end{eqnarray}
where $B(a,b)=\int_{0}^{1}t^{a-1}(1-t)^{b-1}dt$ is the Beta function. We see that there is a topological phase transition $\Delta{\cal C}\neq 0$ if and only if the order of band crossing $n$ is an odd number. Here the extra factor $2nB(\frac{n}{2},n)$ is an artifact of this continuous model, and the jump of topological invariant should be treated as $\Delta{\cal C}=|{\cal C}(M>0)-{\cal C}(M<0)|=n_{+}-n_{-}\in 2{\mathbb Z}+1$, as demonstrated by the following lattice model
\begin{eqnarray}
&&f=d_{1}-id_{2}=\left(\sin k_{x}-i\sin k_{y}\right)^{n_{-}}\;,\;\;\;g=d_{3}=\sin^{n_{-}}k_{z}\;,
\nonumber \\
&&d_{5}=M+\cos k_{x}+\cos k_{y}+\cos k_{z}\;.
\end{eqnarray}
The resulting topological invariant is 
\begin{eqnarray}
{\cal C}&=&\frac{\pi}{3}\int_{BZ}\frac{d^{3}{\bf k}}{(2\pi)^{3}}\epsilon^{\mu\nu\rho}{\rm Tr}\left[q^{\dag}\partial_{\mu}qq^{\dag}\partial_{\nu}qq^{\dag}\partial_{\rho}q\right]
\nonumber \\
&=&\left\{
\begin{array}{ll}
n_{-} & {\rm for}\;M\apprge -3\;{\rm and}\;n_{-}\in 2{\mathbb Z}+1\;, \\
0 & {\rm for}\;M\apprle -3\;{\rm or}\;n_{-}\in 2{\mathbb Z}\;, \\
\end{array}
\right.
\label{3D_class_AIII_topo_inv_BZ}
\end{eqnarray}
which gives $\Delta{\cal C}=n_{-}\in 2{\mathbb Z}+1$.

\begin{figure}
\centering
\includegraphics[width=0.7\linewidth]{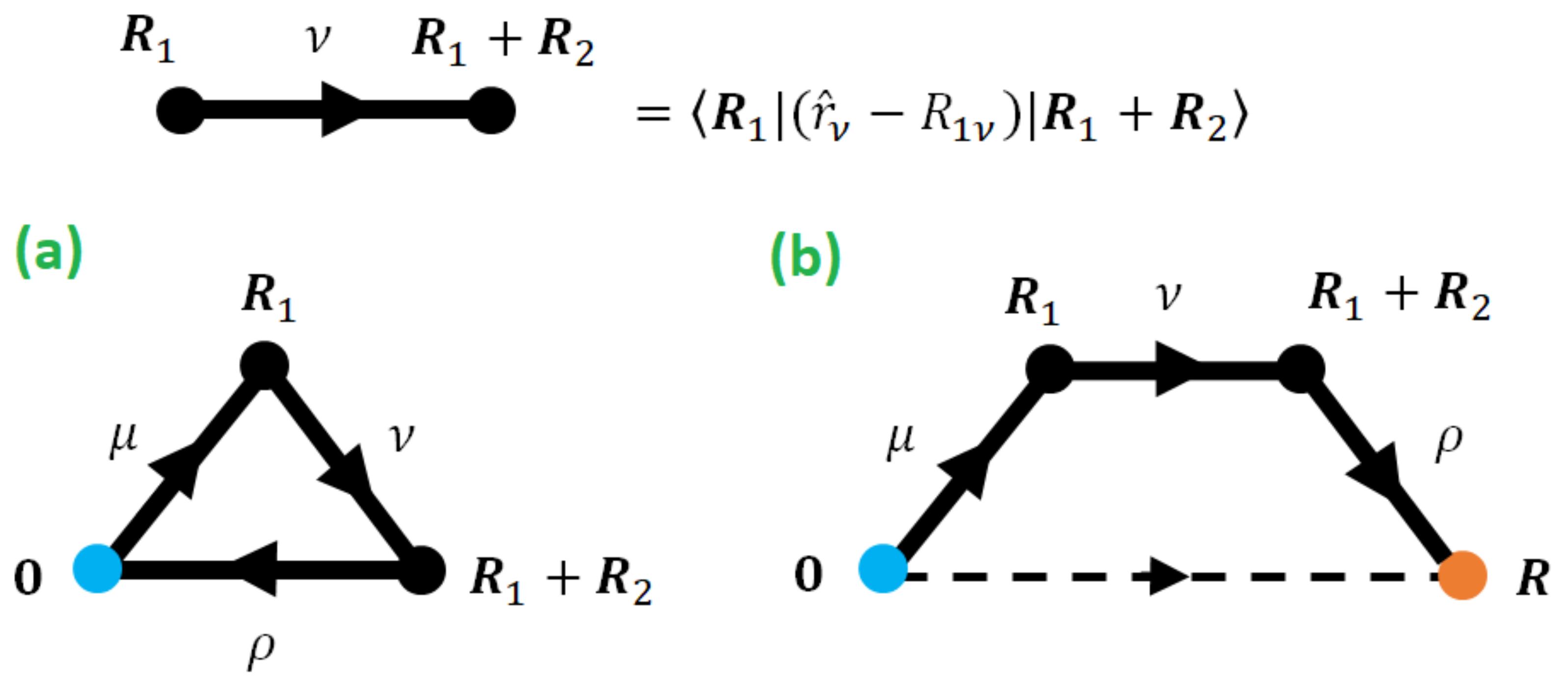}
\caption{Diagrammatic representation of (a) the topological invariant expressed in real space in terms of the Wannier states, and (b) the Wannier state correlation function $\tilde{F}_{3D}({\bf R})$ in 3D class AIII. The two black points ${\bf R}_{1}$ and ${\bf R}_{1}+{\bf R}_{2}$ scan through the entire system and are to be integrated.
}
\label{fig:3D_class_AIII_topo_inv_Wannier}
\end{figure}

The Wannier state correlation function in this class is introduced from the Chern-Simons gauge field ${\cal A}_{\mu}^{ab}=\langle u_{a-}|\partial_{\mu}|u_{b-}\rangle$ defined from the eigenstates for the filled bands $E_{-}({\bf k})=-d$ 
\begin{eqnarray}
|u_{1-}\rangle=\frac{1}{\sqrt{2}d}\left(
\begin{array}{c}
-d_{1}+id_{2} \\
d_{3}+iM \\
0 \\
d
\end{array}
\right),\;\;\;
|u_{2-}\rangle=\frac{1}{\sqrt{2}d}\left(
\begin{array}{c}
-d_{3}+iM \\
-d_{1}-id_{2} \\
d \\
0
\end{array}
\right).
\label{3D_class_AIII_eigenstates}
\end{eqnarray}
We find that the integrand in Eq.~(\ref{3D_class_AIII_topo_inv}) can be written as
\begin{eqnarray}
&&\epsilon^{\mu\nu\rho}{\rm Tr}\left[q^{\dag}\partial_{\mu}qq^{\dag}\partial_{\nu}qq^{\dag}\partial_{\rho}q\right]
=8\epsilon^{\mu\nu\rho}{\rm Tr}\left[{\cal A}_{\mu}{\cal A}_{\nu}{\cal A}_{\rho}\right]
\nonumber \\
&&=\frac{12M}{d^{4}}\epsilon^{\mu\nu\rho}
\partial_{x}d_{\mu}\partial_{y}d_{\nu}\partial_{z}d_{\rho}\;.
\label{3D_class_AIII_Trqdq_AAA_equivalence}
\end{eqnarray}
As a result, the topological invariant in Eq.~(\ref{3D_class_AIII_topo_inv_BZ}) is equivalently 
\begin{eqnarray}
{\cal C}=\frac{8\pi}{3}\int_{BZ}\frac{d^{3}{\bf k}}{(2\pi)^{3}}\epsilon^{\mu\nu\rho}{\rm Tr}\left[{\cal A}_{\mu}{\cal A}_{\nu}{\cal A}_{\rho}\right]\;.
\label{3D_class_AIII_topo_inv_AAA}
\end{eqnarray}
In addition, the Chern-Simons gauge field ${\cal A}_{\mu}^{ab}$ can be expressed as the Fourier transform of the filled-band Wannier states 
\begin{eqnarray}
{\cal A}_{\mu}^{ab}=-i\sum_{\bf R}e^{i{\bf k\cdot R}}\langle{\bf 0}a|{\hat r}_{\mu}|{\bf R}b\rangle=-i\sum_{\bf R}e^{-i{\bf k\cdot R}}\langle{\bf R}a|{\hat r}_{\mu}|{\bf 0}b\rangle\;,
\label{3D_class_AIII_Chern_Simons_gauge_field_Wannier}
\end{eqnarray}
where $|{\bf R}b\rangle$ is the Wannier state of filled band $b$ localized at home cell ${\bf R}$. We use following compact notation for the eigenstates and the corresponding Wannier states
\begin{eqnarray}
|u_{-}\rangle=\left(
\begin{array}{c}
|u_{1-}\rangle \\
|u_{2-}\rangle
\end{array}
\right)\;,\;\;\;
|{\bf R}\rangle=\left(
\begin{array}{c}
|{\bf R}1\rangle \\
|{\bf R}2\rangle
\end{array}
\right)\;.
\end{eqnarray}
It then follows that the topological invariant in Eq.~(\ref{3D_class_AIII_topo_inv_AAA}) can be written in terms of the Wannier states entirely in real space
\begin{eqnarray}
&&{\cal C}=i\frac{8\pi}{3}\int d^{3}{\bf R}_{1}\int d^{3}{\bf R}_{2}\epsilon^{\mu\nu\rho}{\rm Tr}\left[\langle{\bf 0}|{\hat r}_{\mu}|{\bf R}_{1}\rangle\langle{\bf R}_{1}|({\hat r}_{\nu}-R_{1\nu})|{\bf R}_{1}+{\bf R}_{2}\rangle\right.
\nonumber \\
&&\left.\times\langle{\bf R}_{1}+{\bf R}_{2}|({\hat r}_{\rho}-R_{1\rho}-R_{2\rho})|{\bf 0}\rangle\right]\;,
\end{eqnarray}
where we have converted the discrete sum of Bravais points into an integral.
Similarly, we introduce the Wannier state correlation function from the Fourier transform of the integrand in the topological invariant
\begin{eqnarray}
&&\tilde{F}_{3D}({\bf R})=\tilde{F}_{3D}(-{\bf R})=\int_{BZ}\frac{d^{3}{\bf k}}{(2\pi)^{3}}e^{-i{\bf k\cdot R}}\epsilon^{\mu\nu\rho}{\rm Tr}\left[{\cal A}_{\mu}{\cal A}_{\nu}{\cal A}_{\rho}\right]
\nonumber \\
&&=i\int d^{3}{\bf R}_{1}\int d^{3}{\bf R}_{2}\epsilon^{\mu\nu\rho}{\rm Tr}\left[\langle{\bf 0}|{\hat r}_{\mu}|{\bf R}_{1}\rangle\langle{\bf R}_{1}|({\hat r}_{\nu}-R_{1\nu})|{\bf R}_{1}+{\bf R}_{2}\rangle\right.
\nonumber \\
&&\left.\times\langle{\bf R}_{1}+{\bf R}_{2}|({\hat r}_{\rho}-R_{1\rho}-R_{2\rho})|{\bf R}\rangle\right]\;.
\label{3D_class_AIII_correlation_fn}
\end{eqnarray}
These real space representations are presented diagrammatically in Fig.~\ref{fig:3D_class_AIII_topo_inv_Wannier}. In addition, Eq.~(\ref{3D_class_AIII_Trqdq_AAA_equivalence}) and Eq.~(\ref{3D_class_AIII_Trqdq_spherical}) indicates that the curvature function satisfies the scaling form in Eq.~(\ref{general_FkM_scaling_form}) with $D=3$, and hence the correlation length $\xi$ has critical exponent $\nu=1/n$. Finally, the Chern-Simons invariant\cite{Ryu10}, equivalently the magnetoelectric polarizability, can as well be expressed in terms of Wannier states 
\begin{eqnarray}
&&CS_{3}=-\pi\int\frac{d^{3}{\bf k}}{(2\pi)^{3}}\epsilon^{\mu\nu\rho}{\rm Tr}\left[A_{\mu}\partial_{\nu}A_{\rho}+\frac{2}{3}A_{\mu}A_{\nu}A_{\rho}\right]
\nonumber \\
&&=i\pi\int d^{3}{\bf R}_{1}\epsilon^{\mu\nu\rho}{\rm Tr}\left[\langle{\bf 0}|{\hat r}_{\mu}|{\bf R}_{1}\rangle\langle{\bf R}_{1}|R_{1\nu}({\hat r}_{\rho}-R_{1\rho})|{\bf 0}\rangle\right]
\nonumber \\
&&-i\frac{2\pi}{3}\int d^{3}{\bf R}_{1}\int d^{3}{\bf R}_{2}\epsilon^{\mu\nu\rho}{\rm Tr}\left[\langle{\bf 0}|{\hat r}_{\mu}|{\bf R}_{1}\rangle\langle{\bf R}_{1}|({\hat r}_{\nu}-R_{1\nu})|{\bf R}_{1}+{\bf R}_{2}\rangle\right.
\nonumber \\
&&\left.\times\langle{\bf R}_{1}+{\bf R}_{2}|({\hat r}_{\rho}-R_{1\rho}-R_{2\rho})|{\bf R}\rangle\right]\;.
\end{eqnarray}
according to the same argument.

\subsection{3D class DIII \label{sec:3D_class_DIII}}

For 3D class DIII that has $T^{2}=-1$ and $C^{2}=1$, we follow Ref.~\cite{Ryu10} and use the same $\Gamma$-matrices as in Eq.~(\ref{3D_class_AIII_gamma_matrices}). The TR and PH operators are $T=\sigma_{y}\otimes\tau_{x}K$ and $C=\sigma_{y}\otimes\tau_{y}K$. It follows that the TR symmetry in Eq.~(\ref{General_symmetries_TR_PH_CH}) requires 
\begin{eqnarray}
d_{i}({\bf k})=-d_{i}(-{\bf k})\;\;\;{\rm for}\;i=1,2,3,4,\;\;\;d_{5}({\bf k})=d_{5}(-{\bf k})\;.
\label{3D_class_DIII_d_oddness_TR}
\end{eqnarray}
Thus only the $d_{5}$ term can be the mass term $d_{5}=M$. On the other hand, the PH symmetry requires
\begin{eqnarray}
d_{i}({\bf k})=-d_{i}(-{\bf k})\;\;\;{\rm for}\;i=1,2,3,\;\;\;d_{i}({\bf k})=d_{i}(-{\bf k})\;\;\;{\rm for}\;i=4,5.
\label{3D_class_DIII_d_oddness_PH}
\end{eqnarray}
Thus the $d_{4}$ term cannot exist, and the Hamiltonian takes the same form as Eq.~(\ref{3D_class_AIII_Hamiltonian_general}), with the same parametrization on $f$ and $g$ as in Eq.~(\ref{3D_class_AIII_f_g}). Given the rotational symmetry operator in Eq.~(\ref{3D_class_AIII_Cm}), demanding $\left[T,C_{m}\right]=0$ and $\left[C,C_{m}\right]=0$ requires $\alpha_{p}=\alpha_{s}^{\ast}$, $\alpha_{q}=\alpha_{r}^{\ast}$. 

%Combining TR and PH, we see that $\alpha_{p}=\alpha_{q}=\alpha_{r}=\alpha_{s}=1$, which can never be satisfied given that $\alpha_{p}=e^{i\frac{2\pi}{m}\left(p+\frac{1}{2}\right)}$. Thus we conclude that $m$-fold rotational symmetry is not compatible with 3D class DIII systems. 

The rotational symmetry in Eq.~(\ref{general_rotational_symmetry}) requires that  
\begin{eqnarray}
\alpha_{p}^{2}f(k_{\pm},k_{z})=\left(\alpha_{q}^{\ast}\right)^{2}f(k_{\pm},k_{z})
=f(k_{\pm}e^{\pm i\frac{2\pi}{m}},k_{z})\;,
\nonumber \\
\alpha_{p}\alpha_{q}g(k_{\pm},k_{z})=\alpha_{p}^{\ast}\alpha_{q}^{\ast}g(k_{\pm},k_{z})
=g(k_{\pm}e^{\pm i\frac{2\pi}{m}},k_{z})\;,
\label{3D_class_DIII_f_g_constraints}
\end{eqnarray}
from which we see that $\alpha_{p}=\alpha_{q}^{\ast}$, and hence the rotational operator is constrained to be the of form 
\begin{eqnarray}
C_{m}={\rm diag}(\alpha_{p},\alpha_{p}^{\ast},\alpha_{p},\alpha_{p}^{\ast})\;.
\label{3D_class_DIII_Cm_final_form}
\end{eqnarray}
The constraint on $f$ in Eq.~(\ref{3D_class_DIII_f_g_constraints}) renders 
\begin{eqnarray}
e^{i\frac{2\pi}{m}(2p+F)}k_{+}^{n_{+}}k_{-}^{n_{-}}=e^{i\frac{2\pi}{m}(n_{+}-n_{-})}k_{+}^{n_{+}}k_{-}^{n_{-}}\;,
\end{eqnarray}
which together with the requirement that $f=d_{1}-id_{2}$ must be odd in momentum yields
\begin{eqnarray}
n_{+}-n_{-}=xm+2p+F\in 2{\mathbb Z}+1\;,
\label{3D_class_DIII_npnm_p}
\end{eqnarray}
We see that the fermionic case $F=1$ can be satisfied by any of the $m=2,3,4,6$, but bosonic case $F=0$ can only be realized at $m=3$. The topological invariant in this class is the 3D winding number discussed in Sec.~\ref{sec:3D_class_AIII}, and hence the formalism of Wannier state correlation function, correlation length, and critical exponents follows that from Eq.~(\ref{3D_class_AIII_eigenstates}) to Eq.~(\ref{3D_class_AIII_correlation_fn}).

\subsection{3D class AII \label{sec:3D_class_AII}}

For 3D class AII that has $T^{2}=-1$, we use the same representation of the $\Gamma$-matrices as in Eq.~(\ref{3D_class_AIII_gamma_matrices}), and the TR operator $T=i\sigma_{y}\otimes IK$ following Ref.~\cite{Ryu10}.
The TR symmetry requires that 
\begin{eqnarray}
d_{i}({\bf k})=-d_{i}(-{\bf k})\;{\rm for}\;i=1,2,3,5\;,\;\;\;d_{4}({\bf k})=d_{4}(-{\bf k})\;,
\end{eqnarray}
and hence only $d_{4}=M$ can be the mass term, and all others must be odd functions of momentum. Demanding the rotational operator in Eq.~(\ref{3D_class_AIII_Cm}) to commute with the TR operator $\left[C_{m},T\right]=0$, renders $\alpha_{p}=\alpha_{q}^{\ast}$ and $\alpha_{r}=\alpha_{s}^{\ast}$, and consequently $C_{m}={\rm diag}(\alpha_{p},\alpha_{p}^{\ast},\alpha_{r},\alpha_{r}^{\ast})$. We may write the Hamiltonian into the form 
\begin{eqnarray}
H=\left(
\begin{array}{cccc}
M & & g & f \\
 & M & f^{\ast} & -g^{\ast} \\
g^{\ast} & f & -M & \\
f^{\ast} & -g & & -M 
\end{array}
\right)\;,
\end{eqnarray}
where $f=d_{1}-id_{2}$ and $g=d_{3}-id_{5}$. The rotational symmetry in Eq.~(\ref{general_rotational_symmetry}) demands that 
\begin{eqnarray}
&&\alpha_{p}\alpha_{r}f(k_{\pm},k_{z})=f(k_{\pm}e^{\pm i\frac{2\pi}{m}},k_{z})\;,
\nonumber \\
&&\alpha_{p}\alpha_{r}^{\ast}g(k_{\pm},k_{z})=g(k_{\pm}e^{\pm i\frac{2\pi}{m}},k_{z})\;.
\label{3D_class_AII_f_g_constraint}
\end{eqnarray}
We parametrize $f$ and $g$ by  
\begin{eqnarray}
f=k_{+}^{n_{+}}k_{-}^{n_{-}}\;,\;\;\;g=k_{z}^{n}\;.
\end{eqnarray}
With this parametrization in mind, the constraint on $g$ in Eq.~(\ref{3D_class_AII_f_g_constraint}) demands that $p=r$, thus the rotational operator takes the form of Eq.~(\ref{3D_class_DIII_Cm_final_form}). Consequently, we arrive at the same constraint to the rotational eigenvalues as Eq.~(\ref{3D_class_DIII_npnm_p}).

The ${\mathbb Z}_{2}$ invariant ${\cal C}$ for 3D lattice models in class AII can be constructed from the Pfaffian of the filled-band $m$-matrix $m_{\alpha\beta}({\bf k})=\langle u_{\bf \alpha k}|T|u_{\bf \beta k}\rangle$ at HSPs\cite{Fu07,Fu07_2}, provided the Pfaffian is a real number\cite{Fu07,Fu07_2}
\begin{eqnarray}
\left(-1\right)^{\cal C}=\prod_{i}{\rm Sgn}\left({\rm Pf}[m({\bf k}_{0,i})]\right)\; ,
\label{Z2_index_from_sgn_Pfaffian}
\end{eqnarray}
where ${\bf k}_{0,i}$ is the $i$-th HSP. The energy eigenstates for the filled bands $E_{-}({\bf k})=-d$ are 
\begin{eqnarray}
|u_{1-}\rangle=\frac{1}{\sqrt{2d(d+M)}}\left(
\begin{array}{c}
-d_{1}+id_{2} \\
d_{3}+id_{5} \\
0 \\
d+M
\end{array}
\right),
\nonumber \\
|u_{2-}\rangle=\frac{1}{\sqrt{2d(d+M)}}\left(
\begin{array}{c}
-d_{3}+id_{5} \\
-d_{1}-id_{2} \\
d+M \\
0
\end{array}
\right).
\label{3D_class_AII_eigenstates}
\end{eqnarray}
The Pfaffian that enters the $\mathbb{Z}_{2}$ topological invariant in Eq.~(\ref{Z2_index_from_sgn_Pfaffian}) is
\begin{eqnarray}
{\rm Pf}\left[m({\bf k}_{0})\right]=\langle u_{1-}|T|u_{2-}\rangle=-\frac{M}{d}\;.
\end{eqnarray}
The critical behavior of the Pfaffian is that as $M\rightarrow M_{c}$, although the Pfaffian at ${\bf k}_{0}$ remains at $\pm 1$, its second derivative diverges (or equivalently the Laplacian diverges since the model is isotropic around each ${\bf k}_{0}$). This motivates us to consider an additional topological invariant ${\cal C}^{\prime}$ that uses the Laplacian of the Pfaffian as the curvature function
\begin{eqnarray}
F({\bf k},M)=\nabla_{\bf k}^{2}{\rm Pf}[m({\bf k},M)]=\sum_{i=1}^{D}\partial_{i}^{2}{\rm Pf}[m({\bf k},M)]\;.
\label{3D_class_AII_scaling_fn_2nd_derivative}
\end{eqnarray}
This choice of curvature function is sound because it integrates to a topological invariant, which can be seen by considering the integration over the BZ of a $D$-dimensional cubic lattice
\begin{eqnarray}
{\cal C}^{\prime}=\int \frac{d^{D}{\bf k}}{(2\pi)^{D}}\nabla_{\bf k}^{2}{\rm Pf}[m]
=\sum_{j=1}^{D}\partial_{j}\left.\left\{\prod_{i\neq j}\int_{-\pi}^{\pi}\frac{dk_{i}}{2\pi}{\rm Pf}[m]\right\}\right|_{k_{j}=-\pi}^{k_{j}=\pi}=0\;.\;\;\;
\label{C_prime}
\end{eqnarray}
Moreover, since ${\rm Pf}[m]$ is usually an even function, its Laplacian $\nabla_{\bf k}^{2}{\rm Pf}[m]$ is also an even function. Note that since ${\cal C}^{\prime}$ is always zero, it is not directly related to the ${\mathbb Z}_{2}$ invariant in Eq.~(\ref{Z2_index_from_sgn_Pfaffian}).

We introduce the Wannier state correlation function for TR-invariant systems by considering the Fourier transform of the curvature function in Eq.~(\ref{3D_class_AII_scaling_fn_2nd_derivative}). In terms of the Wannier state $|{\bf R}n\rangle$ of the $n$-th filled-band localized at home cell ${\bf R}$, introduced in Eq.~(\ref{Wannier_basis}), the derivatives of the Pfaffian take the form
\begin{eqnarray}
&&\partial_{i}{\rm Pf}[m({\bf k})]=-i\sum_{\bf R}e^{-i{\bf k}\cdot{\bf R}}\langle{\bf R}1|R_{i}T|{\bf 0}2\rangle\;,
\nonumber \\
&&\partial_{i}^{2}{\rm Pf}[m({\bf k})]=-\sum_{\bf R}e^{-i{\bf k}\cdot{\bf R}}\langle{\bf R}1|R_{i}^{2}T|{\bf 0}2\rangle\;.
\end{eqnarray}
The Fourier transform of the Laplacian of the Pfaffian is then
\begin{eqnarray}
\tilde{F}_{TR}({\bf R})=\int \frac{d^{D}{\bf k}}{(2\pi)^{D}}\,e^{i{\bf k}\cdot{\bf R}}{\nabla}_{\bf k}^{2}\,{\rm Pf}[m({\bf k})]=-\langle{\bf R}1|R^{2}\,T|{\bf 0}2\rangle\;,
\label{Wannier_correlation_TR_invariant}
\end{eqnarray}
where $R^{2}=\sum_{i=1}^{D}R_{i}^{2}$. This corresponds to a correlation function that measures the overlap of the Wannier state of the first filled-band centered at ${\bf R}$ and that of the second filled-band centered at the origin, as a matrix element of $R^{2} T$. 

%The same procedure of introducing the Wannier state correlation function follows that in Sec.~\ref{sec:2D_class_AII}, and the Wannier state correlation function is still that in Eq.~(\ref{Wannier_correlation_TR_invariant}). 

\begin{figure}[ht]
\begin{center}
\includegraphics[clip=true,width=0.9\columnwidth]{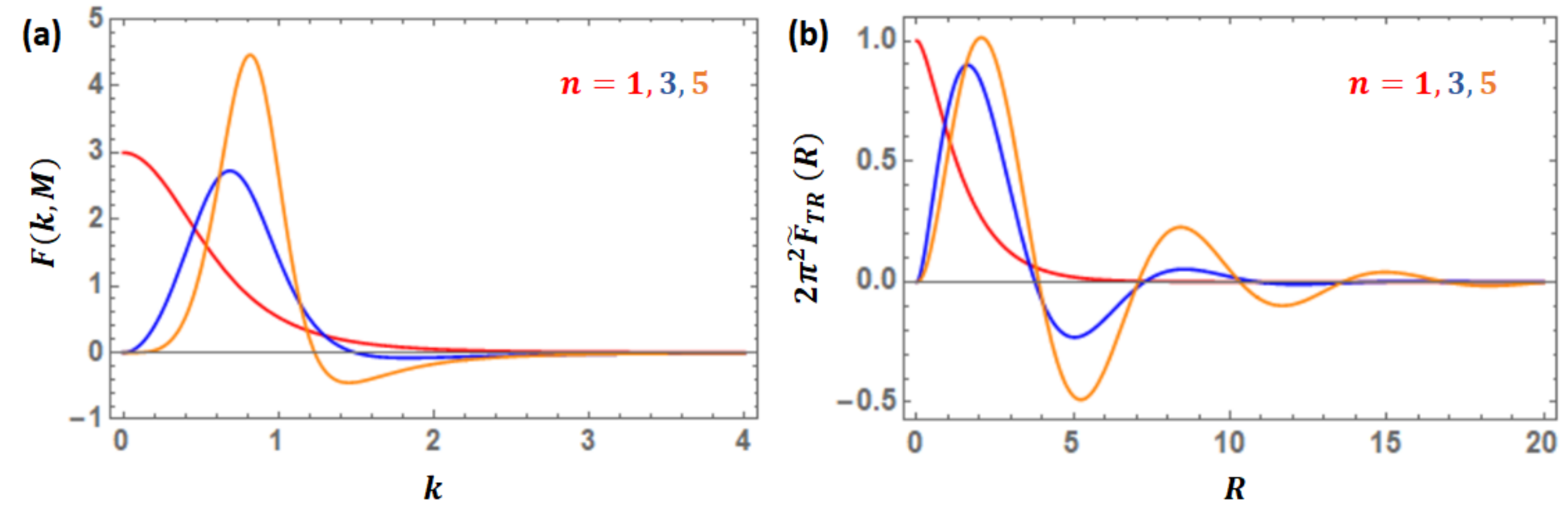}
\caption{ (1) The curvature function and (2) Wannier state correlation function $\tilde{F}_{TR}({\bf R})$ in our continuous model of 3D class AII, at different orders of band crossing. All the curvature functions and correlation functions in the 15 dimension-symmetry classes display a similar behavior. } 
\label{fig:3D_class_AII_corre_fn}
\end{center}
\end{figure}

The Laplacian of the Pfaffian in our continuous model in spherical coordinates is given by 
\begin{eqnarray}
&&{\nabla}_{\bf k}^{2}{\rm Pf}[m]=\frac{1}{k^{2}}\frac{\partial}{\partial k}\left[k^{2}\frac{\partial}{\partial k}\left(-\frac{M}{d}\right)\right]
\nonumber \\
&&=\frac{(n-n^{2})k^{4n-2}M+(2n+1)nk^{2n-2}M^{3}}{\left(k^{2n}+M^{2}\right)^{5/2}}\;,
\end{eqnarray}
whose Fourier transform reads 
\begin{eqnarray}
\tilde{F}_{TR}({\bf R})=\int\frac{d^{3}{\bf k}}{(2\pi)^{3}}e^{i{\bf k\cdot R}}{\nabla}_{\bf k}^{2}{\rm Pf}[m]=\frac{1}{2\pi^{2}R}\int_{0}^{\infty}dk\,k\,\sin kR {\nabla}_{\bf k}^{2}{\rm Pf}[m]\;.
\end{eqnarray}
At any $n$, the correlation function is a decaying function of argument $R/\xi$, with a correlation length $\xi\propto|M|^{-\frac{1}{n}}$ that has critical exponent $\nu=1/n$. At $n=1$, the correlation function monotonically decays, whereas at $n>1$, the correlation function oscillates and decays, as shown in Fig.~\ref{fig:3D_class_AII_corre_fn}. These features remain true for all the correlation functions discussed in the present work.

%{\cblue (1) I think I should also introduce the correlation function from the Fourier transform of the Pfaffian itself. I think I will get $\langle{\bf 0}1|T|{\bf R}2\rangle$, which is easier. Just say alternatively we can as well use this as a correlation function. }

\subsection{3D class CII \label{sec:3D_class_CII}}

For 3D class CII that has $T^{2}=-1$ and $C^{2}=-1$, the minimal model is a $8\times 8$ Dirac model\cite{Schnyder08}. We choose the chiral representation, in which the seven $\Gamma$-matrices are given by\cite{Ryu10} 
\begin{eqnarray}
&&\Gamma^{a}=\Gamma_{4\times 4}^{a}\otimes\eta_{x}\;,\;\;\;{\rm for}\;a=1\sim 4\;,
\nonumber \\
&&\Gamma^{5}=I_{4\times 4}\otimes\eta_{y}\;,\;\;\;\Gamma^{6}=I_{4\times 4}\otimes\eta_{z}\;,
\nonumber \\
&&\Gamma^{7}=(-i)^{3}\Gamma^{1}\Gamma^{2}...\Gamma^{6}\;,
\end{eqnarray}
where $\Gamma_{4\times 4}^{a}$ are those in Eq.~(\ref{3D_class_AIII_gamma_matrices}). $\Gamma^{6}=I_{\sigma}\otimes I_{\tau}\otimes\eta_{z}=S$ is used to implement the chiral symmetry. The Hamiltonian written in the basis of the other six $\Gamma$-matrices is block-off-diagonal
\begin{eqnarray}
&&H({\bf k})=\sum_{i=1,2,3,4,5,7}d_{i}\Gamma^{i}=\left(
\begin{array}{cc}
0 & D \\
D^{\dag} & 0
\end{array}
\right)\;.
\label{3D_class_CII_H_block}
\end{eqnarray}
We will denote $\sigma_{i}$, $\tau_{i}$, and $\eta_{i}$ as the Pauli matrices for the spin, particle-hole, and valley grading. The TR and PH operators may be chosen as $T=i\sigma_{y}K\otimes I_{\tau}\otimes I_{\eta}$ and $C=-i\sigma_{y}K\otimes I_{\tau}\otimes \eta_{z}$. The block-off-diagonal part of the Hamiltonian in Eq.~(\ref{3D_class_CII_H_block}) expressed in terms of the ${\bf d}$-vector yields
\begin{eqnarray}
&&D=\left(
\begin{array}{cccc}
h & & g & f \\
 & h & f^{\ast} & -g^{\ast} \\
g^{\ast} & f & -h^{\ast} & \\
f^{\ast} & -g & & -h^{\ast} 
\end{array}
\right)\;,
\nonumber \\
&&f=d_{1}-id_{2}\;,\;\;\;g=d_{3}+id_{7}\;,\;\;\;h=d_{4}-id_{5}\;.
\end{eqnarray}
The TR symmetry requires that 
\begin{eqnarray}
d_{i}({\bf k})=-d_{i}(-{\bf k})\;\;\;{\rm for}\;i=1,2,3,5,7\;,\;\;\;d_{4}({\bf k})=d_{4}(-{\bf k})\;.
\label{3D_class_CII_di_even_odd}
\end{eqnarray}
Consequently, the mass term is $d_{4}=M$. The PH symmetry also gives the same condition.

We parametrize the $8\times 8$ rotational operator by 
\begin{eqnarray}
C_{m}={\rm diag}(\overline{\alpha}_{p},\overline{\alpha}_{q},\overline{\alpha}_{r},\overline{\alpha}_{s})\;,\;\;\;
\overline{\alpha}_{p}=\left(
\begin{array}{cc}
\alpha_{p_{1}} & \\
 & \alpha_{p_{2}}
\end{array}
\right)\;,
\label{3D_class_CII_Cm_general}
\end{eqnarray}
with $\alpha_{p_{i}}=e^{i\frac{2\pi}{m}\left(p_{i}+\frac{F}{2}\right)}$. The coexistence of TR and rotational symmetry $\left[C_{m},T\right]=0$ requires that $\alpha_{p_{1}}=\alpha_{p_{2}}^{\ast}\equiv \alpha_{p}$. The coexistence of PH and rotational symmetry $\left[C_{m},C\right]=0$ also requires the same thing. As a result, the $\overline{\alpha}_{p}$ in Eq.~(\ref{3D_class_CII_Cm_general}) takes the form $\overline{\alpha}_{p}={\rm diag}(\alpha_{p},\alpha_{p}^{\ast})$.

We now examine the rotational symmetry. To simplify the calculation, we rewrite the Hamiltonian in Eq.~(\ref{3D_class_CII_H_block}) by
\begin{eqnarray}
&&H({\bf k})=\left(
\begin{array}{cccc}
 & & D_{11} & D_{12} \\
 & & D_{21} & D_{22} \\
D_{11}^{\ast} & D_{12} & & \\
D_{21} & D_{22}^{\ast} & & 
\end{array}
\right)\;,
\nonumber \\
&&D_{11}=\left(
\begin{array}{cc}
h & \\
 & h 
\end{array}\right)=-D_{22}^{\ast}\;,\;\;\;
D_{12}=\left(
\begin{array}{cc}
g & f \\
f^{\ast} & -g^{\ast} 
\end{array}\right)=D_{21}^{\dag}\;.
\end{eqnarray}
The rotational symmetry in eq.~(\ref{general_rotational_symmetry}) gives the constraint 
\begin{eqnarray}
&&\overline{\alpha}_{p}D_{12}({\bf k})\overline{\alpha}_{s}^{\ast}=
\overline{\alpha}_{r}D_{12}\overline{\alpha}_{q}^{\ast}=D_{12}(R_{m}{\bf k})\;,
\nonumber \\
&&\overline{\alpha}_{q}D_{21}({\bf k})\overline{\alpha}_{r}^{\ast}=
\overline{\alpha}_{s}D_{21}\overline{\alpha}_{p}^{\ast}=D_{21}(R_{m}{\bf k})\;,
\nonumber \\
&&\overline{\alpha}_{p}D_{11}\overline{\alpha}_{r}^{\ast}=D_{11}(R_{m}{\bf k})\;,
\nonumber \\
&&\overline{\alpha}_{q}D_{22}\overline{\alpha}_{s}^{\ast}=D_{22}(R_{m}{\bf k})\;.
\label{3D_class_CII_Cm_Dij_constraints}
\end{eqnarray}
The first two equations in Eq.~(\ref{3D_class_CII_Cm_Dij_constraints}) yields $\overline{\alpha}_{p}=\overline{\alpha}_{q}=\overline{\alpha}_{r}=\overline{\alpha}_{s}$, so the final form of the rotational operator is 
\begin{eqnarray}
C_{m}={\rm diag}(\overline{\alpha}_{p},\overline{\alpha}_{p},\overline{\alpha}_{p},\overline{\alpha}_{p})\;,\;\;\;
\overline{\alpha}_{p}=\left(
\begin{array}{cc}
\alpha_{p} & \\
 & \alpha_{p}^{\ast}
\end{array}
\right)\;.
\end{eqnarray}
Using the parametrization 
\begin{eqnarray}
f=d_{1}-id_{2}=k_{+}^{n_{+}}k_{-}^{n_{-}}\;,\;\;\;g=d_{3}+id_{7}=d_{3}=k_{z}^{n}\;,\;\;\;
h=d_{4}-id_{5}=d_{4}=M\;,
\nonumber \\
\end{eqnarray}
such that the dispersion satisfies Eq.~(\ref{3D_class_AIII_Ek}), the constraint on $f$ yields 
\begin{eqnarray}
n_{+}-n_{-}=xm+2p+F\in 2{\mathbb Z}+1\;,
\label{3D_class_CII_npnm_pq_constraint}
\end{eqnarray}
because $n=n_{+}+n_{-}\in 2{\mathbb Z}+1$ due to Eq.~(\ref{3D_class_CII_di_even_odd}). For fermions $F=1$, the condition in Eq.~(\ref{3D_class_CII_npnm_pq_constraint}) can be by any of the $m=2,3,4,6$, where as for bosons $F=0$ it can only be satisfied at $m=3$.

%{\cblue (1) Below is Andreas' comment about how to construct the ${\mathbb Z}_{2}$ for class CII: to define the invariant for class CII is quite tricky. In general it is defined by the Chern-Simons invariant, but the wave functions must satisfy certain gauge constraints. That is, one needs to find a suitable basis choice for the occupied states, which is in general quite tricky to find. A general procedure to get this is not known, see the discussion on pages 15 to 19 of my RMP review. Alternatively, one can define the CII invariant by considering an extension to higher dimensions, see discussion on pages 38 to 42 of the NJP paper. Given these difficulties, I think we should be pragmatic and work with the invariant that you have found. It seems to work fine for the considered model. }

%{\cblue (2) Following Andreas' comments above, maybe there is no need to discuss the ${\mathbb Z}_{2}$ invariant here. I can just introduce the correlation function through the Pfaffian without discussing the ${\mathbb Z}_{2}$ invariant. Better not to touch this issue since there might be something that I don't understand, and this is not the purpose of this article anyway. }

We now introduce a winding number and the corresponding Wannier state correlation function. Defining the $4\times 4$ $q$-matrix from the off-diagonal part of the Hamiltonian as\cite{Schnyder08}
\begin{eqnarray}
q({\bf k})=-\frac{1}{d}\left(d_{\mu}\alpha_{\mu}+M\beta\right)=q^{-1}({\bf k})=q^{\dag}({\bf k})\;,
\end{eqnarray}
we found that $\epsilon^{\mu\nu\rho}{\rm Tr}\left[q^{\dag}\partial_{\mu}qq^{\dag}\partial_{\nu}qq^{\dag}\partial_{\rho}q\right]=0$, and hence the simple version of the winding number in Eq.~(\ref{3D_class_AIII_topo_inv}) vanishes. However, if we add a $\Gamma^{5}_{4\times 4}$ (defined in Eq.~(\ref{3D_class_AIII_gamma_matrices})) into the definition 
\begin{eqnarray}
\epsilon^{\mu\nu\rho}{\rm Tr}\left[\Gamma^{5}_{4\times 4}q^{\dag}\partial_{\mu}qq^{\dag}\partial_{\nu}qq^{\dag}\partial_{\rho}q\right]
=24M\frac{k_{\perp}^{2(n-1)}k_{z}^{n-1}(n_{+}-n_{-})n^{2}}{\left(k^{2n}+M^{2}\right)^{2}}\;,
\label{3D_class_CII_Trqdq}
\end{eqnarray}
then it yields a non-vanishing winding number ${\cal C}'$ (see Eq.~(\ref{3D_class_AIII_topo_inv}))
\begin{eqnarray}
{\cal C}'&=&\frac{\pi}{3}\int\frac{d^{3}{\bf k}}{(2\pi)^{3}}\epsilon^{\mu\nu\rho}{\rm Tr}\left[\Gamma^{5}_{4\times 4}q^{\dag}\partial_{\mu}qq^{\dag}\partial_{\nu}qq^{\dag}\partial_{\rho}q\right]
\nonumber \\
&=&\left\{
\begin{array}{ll}
(n_{+}-n_{-}){\rm Sgn}(M)\left[2nB(\frac{n}{2},n)\right] & {\rm if}\;n\in 2{\mathbb Z}+1 \\
0 & {\rm if}\;n\in 2{\mathbb Z}
\end{array}
\right.
\label{3D_class_CII_Cprime}
\end{eqnarray}
where $B(a,b)$ is the Beta function. Although this winding number does not correspond to the true ${\mathbb Z}_{2}$ invariant in this class, the integrand in Eq.~(\ref{3D_class_CII_Cprime}) satisfies all the critical properties of the curvature function discussed in Sec.~\ref{sec:scaling_hypothesis}. In addition, from the eigenstates of the filled bands 
\begin{eqnarray}
|u_{1-}\rangle=\frac{1}{\sqrt{2}d}\left(
\begin{array}{c}
-d_{1}+id_{2} \\
d_{3} \\
0 \\
M \\
0 \\
0 \\
0 \\
d
\end{array}
\right)\;,\;\;\;
|u_{2-}\rangle=\frac{1}{\sqrt{2}d}\left(
\begin{array}{c}
-d_{3} \\
-d_{1}-id_{2} \\
M \\
0 \\
0 \\
0 \\
d \\
0
\end{array}
\right)\;,
\nonumber \\
|u_{3-}\rangle=\frac{1}{\sqrt{2}d}\left(
\begin{array}{c}
0 \\
-M \\
-d_{1}+id_{2} \\
d_{3} \\
0 \\
d \\
0 \\
0
\end{array}
\right)\;,\;\;\;
|u_{4-}\rangle=\frac{1}{\sqrt{2}d}\left(
\begin{array}{c}
-M \\
0 \\
-d_{3} \\
-d_{1}-id_{2} \\
d \\
0 \\
0 \\
0
\end{array}
\right)\;.
\end{eqnarray}
the Berry connection ${\cal A}_{\mu}^{ab}=\langle u_{a-}|\partial_{\mu}|u_{b-}\rangle$ renders 
\begin{eqnarray}
&&\epsilon^{\mu\nu\rho}{\rm Tr}\left[\Gamma^{5}_{4\times 4}q^{\dag}\partial_{\mu}qq^{\dag}\partial_{\nu}qq^{\dag}\partial_{\rho}q\right]
=-8\epsilon^{\mu\nu\rho}{\rm Tr}\left[\Gamma^{5}_{4\times 4}{\cal A}_{\mu}{\cal A}_{\nu}{\cal A}_{\rho}\right]
\nonumber \\
&&=\frac{24M}{d^{4}}\epsilon^{\mu\nu\rho}
\partial_{x}d_{\mu}\partial_{y}d_{\nu}\partial_{z}d_{\rho}\;,
\label{3D_class_CII_Trqdq_AAA_equivalence}
\end{eqnarray}
similar to Eq.~(\ref{3D_class_AIII_Trqdq_AAA_equivalence}). Therefore the formalism of constructing correlation function in Sec.~\ref{sec:3D_class_AIII} can be directly applied to 3D class CII, with the inclusion of $\Gamma_{4\times 4}^{5}$ in the argument of all the traces ${\rm Tr}[...]\rightarrow{\rm Tr}[\Gamma_{4\times 4}^{5}...]$. We denote the resulting correlation function as $\tilde{F}_{3D}^{\prime}({\bf R})$ in Table \ref{tab:universality_class}.

%We then define the true ${\mathbb Z}_{2}$ index ${\cal C}$ by
%\begin{eqnarray}
%{\rm Sgn}({\cal C}')=(-1)^{\cal C}\;,
%\end{eqnarray}
%which is binary at any order of band crossing $n$.

%The matrix $m_{\alpha\beta}=\langle u_{\alpha-}|T|u_{\beta-}\rangle=-m_{\beta\alpha}$ is antisymmetric, and hence its Pfaffian ${\rm Pf}[m]$ is well defined. Direct calculation yields 
%\begin{eqnarray}
%{\rm Pf}[m]=m_{14}m_{23}-m_{13}m_{24}+m_{12}m_{34}=\frac{M^{2}}{d^{2}}=-m_{12}=-m_{34}\;,
%\end{eqnarray}
%indicating that the pair of bands $\left\{1,2\right\}$ should be considered as TR partners, and $\left\{3,4\right\}$ another pair. Following Sec.~\ref{sec:3D_class_AII}, we treat the Laplacian of $m_{12}$ and $m_{34}$ as the curvature function
%\begin{eqnarray}
%{\nabla}_{\bf k}^{2}m_{\alpha\overline{\alpha}}=\frac{(2n-4n^{2})M^{2}k^{4n-2}+2n(2n+1)M^{4}k^{2n-2}}{(k^{2n}+M^{2})^{3}}\;,
%\end{eqnarray}
%whose Fourier transform yields the Wannier state correlation function
%\begin{eqnarray}
%\tilde{F}_{TR}({\bf R})&=&\int\frac{d^{3}{\bf k}}{(2\pi)^{3}}e^{i{\bf k\cdot R}}{\nabla}_{\bf k}^{2}m_{\alpha\overline{\alpha}}=\frac{1}{2\pi^{2}R}\int_{0}^{\infty}dk\,k\,\sin kR\,{\nabla}_{\bf k}^{2}m_{\alpha\overline{\alpha}}
%\nonumber \\
%&=&-\langle{\bf R}\alpha|R^{2}T|{\bf 0}\overline{\alpha}\rangle\;,
%\end{eqnarray}
%where $(\alpha,\overline{\alpha})=(1,2)$ or $(3,4)$ are between TR partners. 

%The correlation function once again decays and oscillates with a correlation length $\xi\sim|M|^{-1/n}$ that indicates the critical exponents $\nu=1/n$. 

\subsection{3D class CI \label{sec:3D_class_CI}}

For 3D class CI that has $T^{2}=1$, $C^{2}=-1$, we use the minimal $8\times 8$ model that writes the Hamiltonian in the chiral basis\cite{Schnyder08} 
\begin{eqnarray}
H=\left(
\begin{array}{cc}
 & D \\
D^{\dag} & 
\end{array}
\right)\;,\;\;\;D({\bf k})=i\sigma_{y}\beta\left(d_{i}\alpha_{i}-id_{4}\gamma^{5}\right)=D^{T}(-{\bf k})\;,
\label{3D_class_CI_H}
\end{eqnarray}
where the $4\times 4$ matrices $\left\{\alpha_{i},\beta,\gamma^{5}\right\}$ are those in Eq.~(\ref{3D_class_AIII_gamma_matrices}). The $D$ matrix expressed in terms of the ${\bf d}=(d_{1},d_{2},d_{3},d_{4})$ vector is 
\begin{eqnarray}
D&=&\left(
\begin{array}{cc}
 & D_{12} \\
D_{21} & 
\end{array}
\right)
=\left(
\begin{array}{cccc}
 & & f^{\ast} & -g \\
 & & -g^{\ast} & -f \\
-f^{\ast} & g & & \\
g^{\ast} & f & &
\end{array}
\right)\;,
\label{3D_class_CI_D}
\end{eqnarray}
with $f=d_{1}-id_{2}$ and $g=d_{3}+id_{4}$. The $8\times 8$ TR and PH operators are $T=I\otimes I\otimes \sigma_{x}K$ and $C=I\otimes I\otimes(-i\sigma_{y})K$.
The TR and PH symmetry require 
\begin{eqnarray}
d_{i}({\bf k})=-d_{i}(-{\bf k})\;\;\;{\rm for}\;i=1,2,3,\;\;\;
d_{4}({\bf k})=d_{4}(-{\bf k})\;.
\label{3D_class_CI_di_oddness}
\end{eqnarray}
Consequently, only $d_{4}=M$ can be the mass term.

Now we consider the $8\times 8$ rotational operator in Eq.~(\ref{3D_class_CII_Cm_general}). The coexistence of TR and rotational symmetry, and the coexistence of PH and rotational symmetry imply $\overline{\alpha}_{p}=\overline{\alpha}_{r}^{\ast}$ and $\overline{\alpha}_{q}=\overline{\alpha}_{s}^{\ast}$, so the rotational operator is constrained to take the form
\begin{eqnarray}
C_{m}={\rm diag}(\overline{\alpha}_{p},\overline{\alpha}_{q},\overline{\alpha}_{p}^{\ast},\overline{\alpha}_{q}^{\ast})\;.
\end{eqnarray}
We use Eqs.~(\ref{3D_class_CI_H}) and (\ref{3D_class_CI_D}) to write the Hamiltonian into the form 
\begin{eqnarray}
H({\bf k})=\left(
\begin{array}{cccc}
 & & & D_{12}({\bf k}) \\
 & & -D_{12}({\bf k}) & \\
 & -D_{12}^{\dag}({\bf k}) & & \\
D_{12}^{\dag}({\bf k}) & & & 
\end{array}
\right)\;,
\end{eqnarray}
Writing down each component in the $8\times 8$ matrix explicitly and using the definition in Eq.~(\ref{3D_class_CII_Cm_general}), we see that rotational symmetry in Eq.~(\ref{general_rotational_symmetry}) requires 
\begin{eqnarray}
&&\alpha_{p_{1}}f^{\ast}({\bf k})\alpha_{q_{1}}=\alpha_{p_{2}}^{\ast}f^{\ast}({\bf k})\alpha_{q_{2}}^{\ast}=f^{\ast}(R_{m}{\bf k})\;,
\nonumber \\
&&\alpha_{p_{2}}g({\bf k})\alpha_{q_{1}}=\alpha_{p_{1}}^{\ast}g({\bf k})\alpha_{q_{2}}^{\ast}=\alpha_{p_{1}}g({\bf k})\alpha_{q_{2}}=g(R_{m}{\bf k})\;.
\end{eqnarray}
To satisfy the Dirac form of the dispersion, we proceed to parametrize 
\begin{eqnarray}
f=d_{1}-id_{2}=k_{+}^{n_{+}}k_{-}^{n_{-}}\;,\;\;\;g=d_{3}+iM=k_{z}^{n}+iM\;,
\end{eqnarray}
with $n=n_{+}+n_{-}\in 2{\mathbb Z}+1$ due to Eq.~(\ref{3D_class_CI_di_oddness}). Since $g$ is invariant under rotation $g({\bf k})=g(R_{m}{\bf k})$, it requires that $\alpha_{p_{2}}\alpha_{q_{1}}=\alpha_{p_{1}}^{\ast}\alpha_{q_{2}}^{\ast}=\alpha_{p_{1}}\alpha_{q_{2}}=1$. A detailed analysis shows that these conditions cannot be satisfied at any nonzero $m$ for either fermions $F=1$ or bosons $F=0$, hence 3D class CI is not compatible with the $m$-fold rotational symmetry.

We proceed to discuss higher-order Dirac model $n\geq 1$ without considering the rotational symmetry. Defining the $q$-matrix
\begin{eqnarray}
q({\bf k})=-\frac{D({\bf k})}{d}\;,\;\;\;q^{\dag}({\bf k})=q^{-1}({\bf k})=-\frac{D^{\dag}({\bf k})}{d}\;,
\end{eqnarray}
the winding number is calculated from the integration of 
\begin{eqnarray}
\epsilon^{\mu\nu\rho}{\rm Tr}\left[q^{\dag}\partial_{\mu}qq^{\dag}\partial_{\nu}qq^{\dag}\partial_{\rho}q\right]
=-M\frac{24k_{\perp}^{2(n-1)}k_{z}^{n-1}(n_{+}-n_{-})n^{2}}
{\left(k^{2n}+M^{2}\right)^{2}}\;,
\end{eqnarray}
which is twice of that in Eq.~(\ref{3D_class_AIII_Trqdq_cylindrical}) for 3D class AIII. As a result, the winding number defined in Eq.~(\ref{3D_class_AIII_topo_inv}) is always an even number ${\cal C}\in 2{\mathbb Z}$. The notion of correlation function and critical exponents then follows the discussion in Sec.~\ref{sec:3D_class_AIII}.

\section{Topological phase transitions in two dimensions \label{sec:two_dimension}}

Among the five topologically nontrivial symmetry classes in two dimensions, as we address in this section, the topological invariant in class A, C, and D are calculated from the same Berry curvature formula, and so follows the same correlation function. On the other hand, the class AII and DIII are described by the ${\mathbb Z}_{2}$ invariant and the correlation function derived accordingly.

\subsection{2D class A \label{sec:2D_class_A}}

The minimal model for a 2D class A model is described by the Hamiltonian
\begin{eqnarray}
H({\bf k})=f({\bf k})\sigma_{+}+f^{\ast}({\bf k})\sigma_{-}+M\sigma_{z}\;.
\label{2D_class_A_H}
\end{eqnarray}
The rotational operation reads $C_{m}={\rm diag}(\alpha_{p},\alpha_{q})$, and hence the rotational symmetry in Eq.~(\ref{general_rotational_symmetry}) requires
\begin{eqnarray}
\alpha_{p}\alpha_{q}^{\ast}f(k_{\pm})=f\left(k_{\pm}e^{\pm i\frac{2\pi}{m}}\right)\;.
\label{2D_class_A_constraint_on_f}
\end{eqnarray}
The parametrization
\begin{eqnarray}
f(k_{\pm})=k_{+}^{n_{+}}k_{-}^{n_{-}}\;,
\label{2D_class_A_f_parametrization}
\end{eqnarray}
yields
\begin{eqnarray}
n_{+}-n_{-}=xm+(p-q)\in{\mathbb Z}\;,
\label{2D_class_A_npnm_pq}
\end{eqnarray}
since there is no nonspatial symmetry in this class to constrain the order of band crossing, one has $n=n_{+}+n_{-}\in{\mathbb Z}$. This condition can be satisfied by any $m$ for either bosons or fermions.

The topological invariant ${\cal C}$ in 2D class A is given by the integration of the Berry curvature $\Omega(k_{x},k_{y})$. The valance band Berry connection as ${\bf A}=\langle u_{\bf k}|i{\boldsymbol\nabla}_{\bf k}|u_{\bf k}\rangle$, which yields the Berry curvature
\begin{eqnarray}
\Omega(k_{x},k_{y})=\partial_{k_{k}}A_{y}-\partial_{k_{y}}A_{x}
=\frac{1}{2d^{3}}\left\{{\bf d}\cdot\partial_{k_{x}}{\bf d}\times\partial_{k_{y}}{\bf d}\right\}\;.
\label{2D_class_A_Berry_curvature}
\end{eqnarray}
The derivatives may be converted into the polar coordinates by 
\begin{eqnarray}
\partial_{k_{x}}d_{i}=\cos\phi\partial_{k}d_{i}-\frac{\sin\phi}{k}\partial_{\phi}d_{i}\;,\;\;\;
\partial_{k_{y}}d_{i}=\sin\phi\partial_{k}d_{i}+\frac{\cos\phi}{k}\partial_{\phi}d_{i}\;,
\end{eqnarray}
which yields the Berry curvature
\begin{eqnarray}
\Omega(k_{x},k_{y})=\frac{(n_{+}^{2}-n_{-}^{2})Mk^{2n-2}}{2\left[k^{2n}+M^{2}\right]^{3/2}}\;.
\label{2D_class_A_Berry_curvature_npnm}
\end{eqnarray}
The topological invariant is calculated by converting the integration over the BZ into an integration over the entire polar plane
\begin{eqnarray}
{\cal C}&=&\int_{BZ}\frac{d^{2}{\bf k}}{2\pi}\Omega(k_{x},k_{y})\approx\frac{1}{2\pi}\int_{0}^{2\pi}d\phi\int_{0}^{\infty}dk\,k\,\Omega(k_{x},k_{y})
\nonumber \\
&=&\frac{n_{+}-n_{-}}{2}{\rm Sgn}(M)\;.
\label{2D_class_A_topo_inv_npnm}
\end{eqnarray}
The above invariant is normalized such that the change of topological invariant across the topological phase transition $\Delta{\cal C}={\cal C}(M>0)-{\cal C}(M<0)=n_{+}-n_{-}$ is consistent with $\Delta{\cal C}=1$ for the usual linear Dirac model $(n_{+},n_{-})=(1,0)$. 

%From Eq.~(\ref{2D_class_A_topo_inv_npnm}) we see that the topological invariant is essentially determined by the quantum number $n_{+}-n_{-}$, which is related to the symmetry eigenvalues of the rotational operator $\left\{p,q\right\}$ by Eq.~(\ref{2D_class_A_npnm_pq}).

The Fourier transform of the Berry curvature can be expressed in terms of the Wannier states by\cite{Marzari12,Gradhand12,Wang06,Chen17} 
\begin{eqnarray}
&&\tilde{F}_{2D}({\bf R})=\int_{BZ}\frac{d^{2}{\bf k}}{(2\pi)^{2}}e^{i{\bf k\cdot R}}\Omega(k_{x},k_{y})=-i\langle{\bf R}|({\bf R\times{\hat r}})_{z}|{\bf 0}\rangle
\nonumber \\
&&=-i\int d^{2}{\bf r}(R_{x}r_{y}-R_{y}r_{x})W^{\ast}({\bf r-R})W({\bf r})\;,
\end{eqnarray}
which measures the overlap of the Wannier functions centering at ${\bf R}$ and at the origin ${\bf 0}$, sandwiched by the factor $R_{x}r_{y}-R_{y}r_{x}$. Within our continuous model formalism
\begin{eqnarray}
\tilde{F}_{2D}({\bf R})\approx\frac{1}{2\pi}\int_{0}^{\infty} dk\,k\,J_{0}(Rk)\Omega(k_{x},k_{y})\;,
\end{eqnarray}
where $J_{0}(Rk)$ is the $0^{\rm th}$-order Bessel function. Using the Berry curvature in Eq.~(\ref{2D_class_A_Berry_curvature_npnm}), we find that the Fourier transform at any $\left\{n_{+},n_{-}\right\}$ gives a decaying function $\lambda({\bf R})$ with a correlation length proportional to $\xi\propto\left|M\right|^{-1/n}\propto|M|^{-\nu}$. Moreover, because $\left\{n_{+},n_{-}\right\}$ are positive numbers, and hence $n\geq|n_{+}-n_{-}|$, we have the inequality
\begin{eqnarray}
\frac{1}{\nu}\geq|\Delta{\cal C}|\;,
\label{2D_class_A_nu_DC_inequality}
\end{eqnarray}
i.e., the order of band crossing can only be greater than or equal to the jump of topological invariant. Finally, we remark that out of all the correlation functions introduced in the present work, only $\tilde{F}_{2D}({\bf R})$ is gauge invariant (because Berry curvature is gauge invariant) and hence measurable. It can be measured by, for instance, performing a Fourier transform of the Berry curvature measured in the cold atom experiments\cite{Jotzu14,Abanin13,Duca15}, from which the critical behavior discussed above may be verified. 

%which may be viewed as the analog of thermodynamic inequalities in statistical mechanics.

%{\cblue (2) I think we can also calculate edge state for this model and prove that the decay length is the same as the correlation length. Let me think about it. }

\subsection{2D class C \label{sec:2D_class_C}}

%{\cblue (Need to ask Andreas, it is a bit strange that 2D class C can only have even orders of band crossing. ---> The Answer is yes. So no need to worry about this.)}

We now discuss 2D class C that has only particle-hole symmetry $C^{2}=-1$. The minimal model is a $2\times 2$ Dirac model with the PH operator $C=\sigma_{y}K$. Using the rotational operator $C_{m}={\rm diag}(\alpha_{p},\alpha_{q})$, the requirement $[C,C_{m}]=0$ yields $\alpha_{q}=\alpha_{p}^{\ast}$, and hence the rotational operator has to take the form $C_{m}={\rm diag}(\alpha_{p},\alpha_{p}^{\ast})$. On the other hand, the PH symmetry requires $d_{i}({\bf k})=d_{i}(-{\bf k})$, i.e., all the three components are even in momentum. Without loss of generality we choose $d_{3}$ to be the mass term, and parametrize the Hamiltonian by 
\begin{eqnarray}
H({\bf k})=\left(
\begin{array}{cc}
M & f \\
f^{\ast} & -M
\end{array}
\right)\;,\;\;\;
f=d_{1}-id_{2}=k_{+}^{n_{+}}k_{-}^{n_{-}}\;.
\label{2D_class_C_H_f}
\end{eqnarray}
The requirement that $d_{i}({\bf k})$ is even in momentum dictates $n\in 2{\mathbb Z}$, indicating that the order of band crossing can only be even.

The rotational symmetry in Eq.~(\ref{general_rotational_symmetry}) renders
\begin{eqnarray}
(\alpha_{p})^{2}f(k_{\pm})
=f(k_{\pm}e^{\pm i\frac{2\pi}{m}})\;,
\end{eqnarray}
and consequently
\begin{eqnarray}
n_{+}-n_{-}=xm+2p+F\in 2{\mathbb Z}\;,
\label{2D_class_C_npmnm_2p_F}
\end{eqnarray}
must be satisfied. As a result, when the system is fermionic $F=1$, it can only have $m=3$ fold rotational symmetry, whereas the bosonic case $F=0$ can be satisfied with any $m$. 

%{\cblue (1) Is 2D class C only restricted to fermions? Or can it be bosons too? Here I assume that it can be either fermionic $F=1$ or bosonic $F=0$. }

The topological invariant is calculated from the same Chern number as in 2D class A in Sec.~\ref{sec:2D_class_A}. Consequently, the Wannier state correlation function and the critical exponent of the correlation length are introduced exactly in the same way, using the formalism from Eq.~(\ref{2D_class_A_Berry_curvature}) to (\ref{2D_class_A_nu_DC_inequality}). However, it should be emphasized that in 2D class C here the jump of topological invariant $\Delta{\cal C}$, order of band crossing, and inverse of the critical exponent are all even numbers, in contrast to 2D class A in which they can be either even or odd.

%{\cblue (1) If 2D class C can only be fermionic AND of the basis of mixed particle and hole $(c_{\bf k\uparrow},c_{\bf -k\downarrow}^{\dag})$, then we should emphasize that the Wannier state is the Wannier state of Bogoliubov quasiparticles. }

%{\cblue (1) A remark on Chiral d-wave: Andreas and Darrick's model of chiral d-wave has the pairing amplitude (ignoring $k_{z}$ component) $\Delta_{\bf k}=\cos k_{x}-\cos k_{y}+i\sin k_{x}\sin k_{y}\propto (k_{x}+ik_{y})^{2}=k_{+}^{2}=f$, which corresponds to $n_{+}=2$ and $n_{-}=0$ in my formalism. This corresponds to $n_{+}-n_{-}=xm+2p+F=xm+2p+1=2$, which cannot be satisfied by any combination of $(x,m,p)$. So I think this chiral d-wave model cannot satisfy rotational symmetry. }

\subsection{2D class D \label{sec:2D_class_D}}

The calculation of 2D class D, where the PH symmetry satisfies $C^{2}=1$, is similar to that in 2D class C presented in Sec.~\ref{sec:2D_class_C}. We follow the formalism in Ref.~\cite{Ryu10} and use the PH operator $C=\sigma_{x}K$. The PH symmetry requires
\begin{eqnarray}
d_{1}({\bf k})=-d_{1}(-{\bf k})\;,\;\;\;d_{2}({\bf k})=-d_{2}(-{\bf k})\;,\;\;\;
d_{3}({\bf k})=d_{3}(-{\bf k})\;.
\end{eqnarray} 
We thereby choose $d_{3}=M$ as the mass term. Using the Hamiltonian and the lowest order expansion of $f$ in Eq.~(\ref{2D_class_C_H_f}), we see that the order of band crossing is odd $n\in 2{\mathbb Z}+1$.

The commutation of rotational operator $C_{m}={\rm diag}(\alpha_{p},\alpha_{q})$ with the PH operator $[C,C_{m}]=0$ yields $\alpha_{q}=\alpha_{p}^{\ast}$. As a result, the rotational eigenvalues take the form $C_{m}={\rm diag}(\alpha_{p},\alpha_{p}^{\ast})$. Equation (\ref{general_rotational_symmetry})
renders the following condition on $f$
\begin{eqnarray}
(\alpha_{p})^{2}f(k_{\pm})=e^{i\frac{2\pi}{m}(2p+F)}f(k_{\pm})
=f(k_{\pm}e^{\pm i\frac{2\pi}{m}})=k_{+}^{n_{+}}k_{-}^{n_{-}}e^{i\frac{2\pi}{m}(n_{+}-n_{-})}\;.
\end{eqnarray}
Therefore, 
\begin{eqnarray}
n_{+}-n_{-}=xm+2p+F\in 2{\mathbb Z}+1\;,
\label{2D_class_D_npmnm_2p_F}
\end{eqnarray}
must be satisfied. If the system is fermionic $F=1$, this condition can be satisfied in any $m$-fold rotational symmetry, whereas if the system is bosonic $F=0$, then only $m=3$ is allowed.

The topological invariant is calculated from the Chern number in Sec.~\ref{sec:2D_class_A}, and the Wannier state correlation function and the critical exponent of the correlation length are introduced by the formalism from Eq.~(\ref{2D_class_A_Berry_curvature}) to (\ref{2D_class_A_nu_DC_inequality}). However, it should be emphasized that in 2D class D here the jump of topological invariant $\Delta{\cal C}$, order of band crossing, and inverse of the critical exponent are all odd numbers.

\subsection{2D class AII \label{sec:2D_class_AII}}

In this section we discuss 2D class AII that has $T^{2}=-1$, with the consideration of $m$-fold rotational symmetry\cite{Yang14}. The minimal model is a $4\times 4$ Dirac Hamiltonian. We use the representation for the Bernevig-Hughes-Zhang (BHZ) model\cite{Bernevig06,Bernevig13}
\begin{eqnarray}
\Gamma_{a}=\left\{s_{x}\otimes\sigma_{z},s_{y}\otimes I,s_{z}\otimes I,s_{x}\otimes\sigma_{x},s_{x}\otimes\sigma_{y}\right\}\; .
\label{2D_class_AII_Gamma_mat}
\end{eqnarray}
The time-reversal operator in this basis reads $T=-iI\otimes\sigma_{y}K$. The coexistence of TR and rotational symmetry $C_{m}={\rm diag}(\alpha_{p},\alpha_{q},\alpha_{r},\alpha_{s})$ requires $\alpha_{r}=\alpha_{p}^{\ast}$ and $\alpha_{s}=\alpha_{q}^{\ast}$. 
Consequently, $C_{m}={\rm diag}(\alpha_{p},\alpha_{q},\alpha_{p}^{\ast},\alpha_{q}^{\ast})$,
in agreement with Ref.~\cite{Yang14}.

Now the representation in Eq.~(\ref{2D_class_AII_Gamma_mat}) indicates that the Hamiltonian takes the form 
\begin{eqnarray}
H=\left(
\begin{array}{cccc}
M & f & & g \\
f^{\ast} & -M & g^{\ast} & \\
 & g & M & -f^{\ast} \\
g^{\ast} & & -f & -M 
\end{array}
\right)\;,
\label{2D_class_AII_ddotGamma}
\end{eqnarray}
where we have defined $f$ and $g$, and we assume their leading order expansions are
\begin{eqnarray}
f=d_{1}-id_{2}=k_{+}^{n_{+}}k_{-}^{n_{-}}\;,\;\;\;
g=d_{4}-id_{5}=k_{+}^{\ell_{+}}k_{-}^{\ell_{-}}\;,\;\;\;
d_{3}=M\;.
\label{2D_class_AII_f_g_parametrization}
\end{eqnarray}
Note that TR invariance requires
\begin{eqnarray}
d_{i}(-{\bf k})=-d_{i}({\bf k})\;\;\;{\rm for}\;i=1,2,4,5\;,\;\;\;d_{3}({\bf k})=d_{3}(-{\bf k})\;,
\label{2D_class_AII_dkmdmk}
\end{eqnarray}
so we use $d_{3}=M$ as the mass term. The oddness in Eq.~(\ref{2D_class_AII_dkmdmk}) requires the expansion in Eq.~(\ref{2D_class_AII_f_g_parametrization}) to satisfy $n\in 2{\mathbb Z}+1$ and $\ell_{+}+\ell_{-}=2{\mathbb Z}+1$. Equation (\ref{general_rotational_symmetry}) yields the following constraints on $f$ and $g$
\begin{eqnarray}
&&\alpha_{p}\alpha_{q}^{\ast}f(k_{\pm})=e^{i\frac{2\pi}{m}(p-q)}f(k_{\pm})
=f(k_{\pm}e^{\pm i2\pi/m})=k_{+}^{n_{+}}k_{-}^{n_{-}}e^{i\frac{2\pi}{m}(n_{+}-n_{-})}\;,
\nonumber \\
&&\alpha_{p}\alpha_{q}g(k_{\pm})=e^{i\frac{2\pi}{m}(p+q+1)}g(k_{\pm})
=g(k_{\pm}e^{\pm i2\pi/m})=k_{+}^{\ell_{+}}k_{-}^{\ell_{-}}e^{i\frac{2\pi}{m}(\ell_{+}-\ell_{-})}\;,
\nonumber \\
&&\alpha_{p}\alpha_{q}g^{\ast}(k_{\pm})=e^{i\frac{2\pi}{m}(p+q+1)}g^{\ast}(k_{\pm})
=g(k_{\pm}e^{\pm i2\pi/m})=k_{-}^{\ell_{+}}k_{+}^{\ell_{-}}e^{i\frac{2\pi}{m}(\ell_{-}-\ell_{+})}.
\end{eqnarray}
The last two equations of the above equation can never be satisfied at the same time, so we conclude that $g=d_{4}-id_{5}=0$. Thus rotational symmetry requires that $\Gamma_{4}$ and $\Gamma_{5}$ terms cannot exist. The constaint on $f$ then gives 
\begin{eqnarray}
n_{+}-n_{-}=xm+(p-q)\in 2{\mathbb Z}+1\;,
\label{2D_class_AII_npmnq_pq}
\end{eqnarray} 
which can be satisfied by any $m$ for either fermionic or bosonic cases.

Denoting $d=\sqrt{d_{1}^{2}+d_{2}^{2}+d_{3}^{2}}$, the four eigenstates in a gauge choice that makes the Pfaffian real and equal to $\pm 1$ at HSPs are 
\begin{eqnarray} 
|u_{1,3}\rangle=\frac{1}{\sqrt{2d(d\mp d_{3})}}
\left(
\begin{array}{c}
0 \\
0 \\
-d_{3}\pm d \\
d_{1}-id_{2}
\end{array}
\right)\;,
\nonumber \\
|u_{2,4}\rangle=\frac{1}{\sqrt{2d(d\mp d_{3})}}
\left(
\begin{array}{c}
d_{3}\mp d \\
d_{1}+id_{2} \\
0 \\
0
\end{array}
\right)\;,
\label{BHZ_eigenstates}
\end{eqnarray}
where the upper signs are for the occupied states $\left\{1,2\right\}$ which have eigenenergies $E_{1}=E_{2}=-d$, and the lower signs are for the unoccupied states $\left\{3,4\right\}$ which have eigenenergies $E_{3}=E_{4}=d$. The Pfaffian of the $m$-matrix of the two occupied states is
\begin{eqnarray}
{\rm Pf}(m)=m_{12}=\langle u_{1}|T|u_{2}\rangle
=\frac{d_{3}}{d}=\frac{M}{\left(k^{2n}+M^{2}\right)^{1/2}}\;.
\label{2D_class_AII_Pfaffian}
\end{eqnarray}
Thus the sign of $d_{3}/d$ at HSPs determines the topology of the system according to Eq.~(\ref{Z2_index_from_sgn_Pfaffian}).

Using the Pfaffian in Eq.~(\ref{2D_class_AII_Pfaffian}), its Laplacian is 
\begin{eqnarray}
&&\nabla_{\bf k}^{2}{\rm Pf}[m({\bf k})]=\frac{1}{k}\frac{\partial}{\partial k}\left(k\frac{\partial}{\partial k}\frac{d_{3}}{d}\right)+\frac{1}{k^{2}}\frac{\partial^{2}}{\partial\phi^{2}}\frac{d_{3}}{d}
\nonumber \\
&&=\frac{n^{2}Mk^{4n-2}-2n^{2}M^{3}k^{2n-2}}
{\left[k^{2n}+M^{2}\right]^{5/2}}\;.
\label{2D_class_AII_Laplacian_Pfaffian}
\end{eqnarray}
The correlation function in our continuous model renders 
\begin{eqnarray}
\tilde{F}_{TR}({\bf R})\approx\frac{1}{2\pi}\int_0^{\infty}d k\, k J_0(R k)\nabla_{\bf k}^{2}{\rm Pf}[m({\bf k})]\;,
\end{eqnarray}
which displays the same behavior as other correlation functions.
%At any order of band crossing $n$, the correlation function is a decaying function with correlation length $\xi\propto\left|M\right|^{-1/n}$. For $n>1$, the correlation function also oscillates with the same length scale. This indicates the critical exponents $\nu=1/n$, with $n$ an odd integer.

%\begin{figure}
%\centering
%\includegraphics[width=0.9\linewidth]{2D_class_AII_correlation_fn}
%\caption{ (a) The curvature function (Berry curvature) and (b) the corresponding Wannier state correlation function $\tilde{F}_{TR}({\bf R})$ for our continuous model of 2D class AII, at different odd orders of band crossing $n$. The mass term is fixed at $M=1$. }
%\label{fig:2D_class_AII_correlation_fn}
%\end{figure}

\subsection{2D class DIII \label{sec:2D_class_DIII}}

The discussion of 2D class DIII follows that in 3D class DIII with a dimensional reduction $k_{z}=0$, using the same $\Gamma$-matrices and TR and PH operators described in Sec.~\ref{sec:3D_class_DIII}. The analysis from Eq.~(\ref{3D_class_DIII_d_oddness_TR}) to (\ref{3D_class_DIII_Cm_final_form}) remains true in 2D.
%As also discussed in Sec.~\ref{sec:3D_class_DIII}, the $m$-fold rotational symmetry is not compatible with class DIII systems, which is also true in 2D. 
We now translate the 3D Hamiltonian in Eq.~(\ref{3D_class_AIII_Hamiltonian_general}) into 2D and parametrize $f$ and $g'$ by
\begin{eqnarray}
f=k_{+}^{n_{+}}k_{-}^{n_{-}}=d_{1}-id_{2}\;,\;\;\;g'=k_{+}^{\ell_{+}}k_{-}^{\ell_{-}}-iM=d_{3}-iM\;.
\end{eqnarray}
First let us look at the parametrization of $g'$. The constraint on $g'$ is $g'(k_{\pm})=g'(k_{\pm}e^{\pm i\frac{2\pi}{m}})$. If one parametrizes $g'=k_{+}^{\ell_{+}}k_{-}^{\ell_{-}}-iM$, this means $k_{+}^{\ell_{+}}k_{-}^{\ell_{-}}=e^{i\frac{2\pi}{m}(\ell_{+}-\ell_{-})}k_{+}^{\ell_{+}}k_{-}^{\ell_{-}}$, and hence $\ell_{+}=\ell_{-}=\ell/2$, which implies $g=k^{\ell}-iM$ only depends on the module of momentum, which contradicts the requirement that the $d_{3}({\bf k})$ in $g'=d_{3}-iM$ must be odd in momentum. Thus we conclude that the $d_{3}$ term cannot exist, 
% The $d_{3}$ term must be real, and hence it can only be of the form $d_{3}=k_{+}^{\ell_{+}}k_{-}^{\ell_{-}}=k^{\ell}$, but then it will violate the oddness $d_{3}({\bf k})=-d_{3}(-{\bf k})$, and hence the $d_{3}$ term cannot exist 
and we have eventually $g'=-iM$. The oddness of $d_{1}$ and $d_{2}$ again yields the same constraint between the $n_{+}-n_{-}$ and the rotational eigenvalues as in Eq.~(\ref{3D_class_DIII_npnm_p}), which can be satisifed by any $m$ for fermions, and $m=3$ for bosons.
%\begin{eqnarray}
%n_{+}+n_{-}\in 2{\mathbb Z}+1\;,
%\end{eqnarray}
%and hence the order of band crossing $n=n_{+}+n_{-}$ can only be odd.

Using the eigenstates in Eq.~(\ref{3D_class_AIII_eigenstates}), with setting $d_{3}=0$ as discussed above, the ${\mathbb Z}_{2}$ topological invariant can be constructed from the Pfaffian of the $m$-matrix using the formalism in Sec.~\ref{sec:2D_class_AII}. The TR operator in Sec.~\ref{sec:3D_class_DIII} acting on the filled-band eigenstates yields 
\begin{eqnarray}
T|u_{1-}\rangle=\frac{1}{\sqrt{2}d}\left(
\begin{array}{c}
-id \\
0 \\
-M \\
-id_{1}+d_{2}
\end{array}
\right)\;,\;\;\;
T|u_{2-}\rangle=\frac{1}{\sqrt{2}d}\left(
\begin{array}{c}
0 \\
id \\
id_{1}+d_{2} \\
M
\end{array}
\right)\;.
\label{2D_class_DIII_Tu}
\end{eqnarray}
The Pfaffian of the $m$-matrix is then
\begin{eqnarray}
{\rm Pf}\left[m({\bf k})\right]=\langle u_{1-}|T|u_{2-}\rangle=-\langle u_{2-}|T|u_{1-}\rangle=\frac{M}{d}\;.
\label{2D_class_DIII_Pfaffian}
\end{eqnarray}
So we see that the sign of $M$ determines the ${\mathbb Z}_{2}$ index in our continuous model. Obviously, because of the ${\mathbb Z}_{2}$ invariant, we always have $\Delta{\cal C}=1$. The rest of the correlation function, critical exponents, and universality class follows that of Sec.~\ref{sec:2D_class_AII} for 2D class AII.

\section{Topological phase transitions in one dimension \label{sec:one_dimension}}

In one dimension, the topological invariant in class BDI, AIII, and D are described by the same Zak phase. The class DIII is described by the ${\mathbb Z}_{2}$ invariant calculated from the TR operator. On the other hand, the topology of class CII is given by a winding number of its own kind. We will detail all these calculations in this section.

\subsection{1D class BDI \label{sec:1D_class_BDI}}

%{\cblue (1) I think in this class because it has TR symmetry, one may be able to construct correlation function from Pfaffian as well. Check this.}

Following Ref.~\cite{Ryu10}, we model the 1D class BDI systems, which have $T^{2}=1$ and $C^{2}=1$, by $T=\sigma_{z}K$ and $C=\sigma_{x}K$. The TR symmetry requires
\begin{eqnarray}
d_{1}(k)=-d_{1}(-k)\;,\;\;\;d_{2}(k)=d_{2}(-k)\;,\;\;\;d_{3}(k)=d_{3}(-k)\;.
\end{eqnarray}
On the other hand, the PH symmetry requires
\begin{eqnarray}
d_{1}(k)=-d_{1}(-k)\;,\;\;\;d_{2}(k)=-d_{2}(-k)\;,\;\;\;d_{3}(k)=d_{3}(-k)\;.
\end{eqnarray}
Combining the TR and PH symmetry, we see that the $d_{2}$ term cannot exist, and we may use $d_{3}=M$ as the mass term. The oddness of $d_{1}$ implies that we can parametrize the Hamiltonian by
\begin{eqnarray}
H(k)=k^{n}\sigma_{x}+M\sigma_{z}\;,\;\;\;E_{\pm}(k)=\pm\sqrt{k^{2n}+M^{2}}\;,
\label{1D_class_BDI_H_original}
\end{eqnarray}
with an odd order of band crossing $n\in 2{\mathbb Z}+1$.

\begin{figure}
\centering
\includegraphics[width=0.5\linewidth]{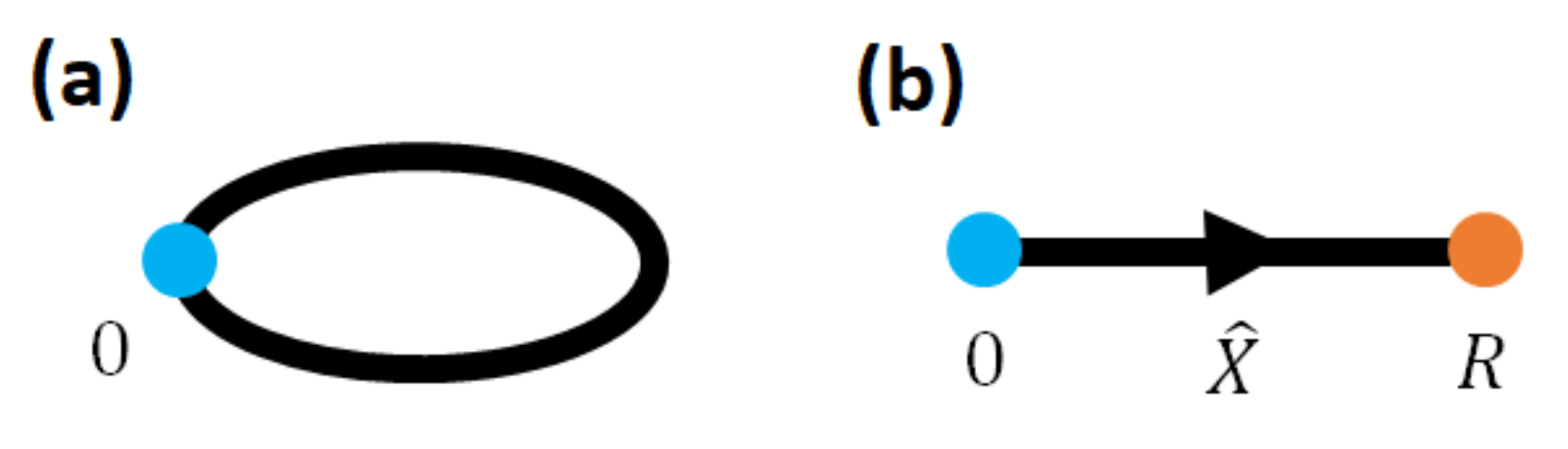}
\caption{ Diagrammatic representation of (a) the topological invariant in real space and (b) the Wannier state correlation function $\tilde{F}_{1D}(R)$ for 1D class BDI, where the position operator is ${\hat X}={\hat r}$. Figure (b) can also be used to represent the correlation functions $\tilde{F}_{TR}({\bf R})$ and $\tilde{F}_{2D}({\bf R})$, where the position operator is replaced by ${\hat X}\rightarrow R^{2}T$ and ${\hat X}\rightarrow({\hat {\bf R}}\times{\hat{\bf r}})_{z}$, respectively.
}
\label{fig:1D_class_BDI_topo_inv_Wannier}
\end{figure}

Denoting the valance band eigenstate of the Hamiltonian in Eq.~(\ref{1D_class_BDI_H_original}) by $|u_{-}\rangle$, the corresponding Berry connection actually vanishes $A(k)=\langle u_{-}|i\partial_{k}|u_{-}\rangle$. Thus for the sake of constructing the topological invariant and introducing the correlation function, we rotate the Hamiltonian by 
\begin{eqnarray}
&&R=e^{i\sigma_{x}\pi/4}=\frac{1}{\sqrt{2}}\left(
\begin{array}{cc}
1 & i \\
i & 1
\end{array}
\right)\;,
\nonumber \\
&&\tilde{H}(k)=RH(k)R^{-1}=d_{1}\sigma_{x}+d_{3}\sigma_{y}=k^{n}\sigma_{x}+M\sigma_{y}\;.
\label{1D_class_BDI_rotate_H}
\end{eqnarray}
That is, we rotate the basis such that the mass term resides in the $\sigma_{y}$ component. The eigenstate in this basis $\tilde{H}(k)|\tilde{u}_{-}\rangle=E(k)|\tilde{u}_{-}\rangle$ is related to the original basis by a rotation $|\tilde{u}_{-}\rangle=V_{k}|u_{-}\rangle$, and gives a Berry connection 
\begin{eqnarray}
\tilde{A}(k)=\langle \tilde{u}_{-}|i\partial_{k}|\tilde{u}_{-}\rangle=
\langle u_{-}|i(V_{k}\partial_{k}V_{k})|u_{-}\rangle=\frac{d_{3}\partial_{k}d_{1}-d_{1}\partial_{k}d_{3}}{2d^{2}}\;.
\end{eqnarray}
Using the parametrization in Eq.~(\ref{1D_class_BDI_H_original}), we obtain the Berry connection
\begin{eqnarray}
\tilde{A}(k)=\frac{nMk^{n-1}}{2(k^{2n}+M^{2})}\;.
\label{1D_class_BDI_Berry_connection}
\end{eqnarray}
The topological invariant is the Zak phase calculated from the integration of Berry connection, which in our continuous model reads
\begin{eqnarray}
{\cal C}=\int_{-\infty}^{\infty}\frac{dk}{\pi}\tilde{A}(k)=\frac{1}{2}{\rm Sgn}(M)\;,
\label{1D_class_BDI_Zak_phase}
\end{eqnarray}
normalized such that jump of topological invariant across the critical point $M_{c}=0$ is $\Delta{\cal C}={\cal C}(M>0)-{\cal C}(M<0)=1$, consistent with the result of the usual linear Dirac model. Following the recipe in Sec.~\ref{sec:curvature_fn_correlation_fn}, we consider our model as an low-energy effective theory of a lattice model, and introduce the Wannier states $|R\rangle$ constructed from $|\tilde{u}_{-}\rangle$. The topological invariant constructed from the Wannier state reads, according to the theory of charge polarization\cite{KingSmith93,Resta94}, 
\begin{eqnarray}
\frac{{\cal C}}{2}=\langle 0|{\hat r}|0\rangle\;.
\end{eqnarray}
The correlation function is given by the Fourier transform of the Berry connection
\begin{eqnarray}
\tilde{F}_{1D}(R)=\int_{BZ}\frac{dk}{2\pi}e^{ikR}\tilde{A}(k)=\langle 0|{\hat r}|R\rangle\;,
\label{1D_class_BDI_correlation_fn}
\end{eqnarray}
which measures the overlap of the Wannier state centering at the origin and that centering at $R$, sandwiched by a position operator ${\hat r}$. These Wannier state representations are shown diagrammatically in Fig.~\ref{fig:1D_class_BDI_topo_inv_Wannier}.

For $n=1$, the Berry connection $\tilde{A}(k)$ displays an extremum $\tilde{A}_{max}$ at the HSP $k=0$, whereas for any $n>1$, the extremum is located at a finite momentum $k_{max}$, i.e., showing a double peak structure in momentum space. 
%The Wannier state correlation function as a Fourier transform of the Berry curvature is an exponentially decayed function at $n=1$, whereas at $n>1$ it is a decayed and oscillating function, with the correlation length proportional to $\xi\propto |M|^{-1/n}$. 
It is straight forward to obtain
\begin{eqnarray}
k_{max}=\left(\frac{n-1}{n+1}M^{2}\right)^{1/2n}\;,\;\;\;
\tilde{A}_{max}=\frac{n}{2}\left(\frac{n-1}{n+1}\right)^{\frac{n-1}{2n}}M^{-1/n}\;.
\end{eqnarray} 
indicating a critical exponent $\alpha=1/n$. Moreover, the half-width-at-half-maximum scales as $k_{HWHM}\propto |M|^{1/n}$, implying the correlation length $\xi\propto |M|^{-1/n}$ and a critical exponent $\nu=1/n$.

%{\cblue (1) Can I put my artificially engineered next-nearest neighbor SSH model here as an example of quadratic band touching in 1D? Ask Andreas to make sure it is still class BDI.  }

\subsection{1D class AIII \label{sec:1D_class_AIII}}

For 1D class AIII that satisfies the chiral symmetry $S^{2}=1$, we follow Ref.~\cite{Ryu10}. The chiral operator is $S=\sigma_{y}$, and hence the chiral symmetry demands that the $d_{2}$ term must vanish. We choose $d_{3}$ to be the mass term and write
\begin{eqnarray}
H(k)=k^{n}\sigma_{x}+M\sigma_{z}\;.
\end{eqnarray}
However, there is no restriction on the order of band crossing $n\in{\mathbb Z}$ being even or odd.

Since the Hamiltonian takes exactly the same form as that for 1D class BDI discussed in Sec.~\ref{sec:1D_class_BDI}, the formalism therein also applies. We make the same rotation of basis by Eq.~(\ref{1D_class_BDI_rotate_H}), and introduce the Berry connection and Zak phase. The resulting Wannier state correlation function takes the same form as in Eq.~(\ref{1D_class_BDI_correlation_fn}). The critical exponent of the correlation length is again $\nu=1/n$, except one should keep in mind that in 1D class AIII here $n$ can be either even or odd.

\subsection{1D class DIII \label{sec:1D_class_DIII}}

The 1D class DIII formalism can be obtained from 3D class DIII in Sec.~\ref{sec:3D_class_DIII} by a dimensional reduction $k_{y}=k_{z}=0$. Since rotational symmetry is absent in 1D, the issue now is reduced to how to parametrize $f=d_{1}-id_{2}$ and $g'=d_{3}-iM$ in Eq.~(\ref{3D_class_AIII_Hamiltonian_general}) using 1D momentum $k_{x}=k$, provided $d_{i}(k)=-d_{i}(-k)$ for $i=1,2,3$. Given the oddness of $d_{i}$, we should parametrize 
\begin{eqnarray}
f=a_{1}k^{n}-ia_{2}k^{n}\;,\;\;\;g'=a_{3}k^{n}-iM\;,\;\;\;n\in 2{\mathbb Z}+1\;,
\end{eqnarray}
such that the dispersion is 
\begin{eqnarray}
E_{\pm}(k)=\pm\sqrt{|f|^{2}+|g'|^{2}}=\pm\left[(a_{1}^{2}+a_{2}^{2}+a_{3}^{2})k^{2n}+M^{2}\right]^{1/2}\;.
\end{eqnarray}
Therefore, the most generalized form is that all $(d_{1},d_{2},d_{3})$ are nonzero, and we denote $d=\sqrt{d_{1}^{2}+d_{2}^{2}+d_{3}^{2}+M^{2}}$. The filled band eigenstates are that in Eq.~(\ref{3D_class_AIII_eigenstates}), since we use the same representation of the $\Gamma$-matrices. Applying the TR operator in Sec.~\ref{sec:3D_class_DIII} to the filled band eigenstates gives the same result as Eq.~(\ref{2D_class_DIII_Tu}), and the Pfaffian is that in Eq.~(\ref{2D_class_DIII_Pfaffian}). So once again the sign of the mass term $M$ determines the ${\mathbb Z}_{2}$ topological invariant in this continuous model.

The Wannier state correlation function in this class is the 1D version of that in Sec.~\ref{sec:2D_class_AII}. The correlation function is again calculated from the Fourier transform of the second derivative of the Pfaffian on the $m$-matrix
\begin{eqnarray}
\partial_{k}^{2}{\rm Pf}\left[m\right]=\frac{(n+n^{2})Mk^{4n-2}+(n-2n^{2})M^{3}k^{2n-2}}{\left(k^{2n}+M^{2}\right)^{5/2}}\;.
\end{eqnarray}
The Fourier transform of which gives 
\begin{eqnarray}
&&\tilde{F}_{TR}(R)=\int_{BZ}\frac{dk}{2\pi}e^{ikR}\partial_{k}^{2}{\rm Pf}[m(k)]=-\langle R1|R^{2}T|02\rangle
\nonumber \\
&&\approx\frac{1}{\pi}\int_{0}^{\infty}dk\,e^{ikR}\partial_{k}^{2}{\rm Pf}[m(k)]\;,
\end{eqnarray}
where $|R1\rangle$ denotes the Wannier state of band $1$ centered at $R$, and $|02\rangle$ denotes the Wannier state of band $2$ centered at the origin. 

%The correlation function decays with a correlation length that scales like $\xi\sim (1/M)^{1/n}$, yielding critical exponent $\nu=1/n$ that is determined by the order of band crossing. 

%The correlation function for several values of $n$ is shown in Fig.~\ref{fig:1D_class_DIII_corre_fn}.

%\begin{figure}
%\centering
%\includegraphics[width=0.5\linewidth]{1D_class_DIII_corre_fn}
%\caption{ Correlation function in 1D class DIII.}
%\label{fig:1D_class_DIII_corre_fn}
%\end{figure}

\subsection{1D class D \label{sec:1D_class_D}}

For 1D class D with $C^{2}=1$, we follow Ref.~\cite{Chiu16} to use the PH operator $C=\sigma_{x}K$, which requires that 
\begin{eqnarray}
d_{i}(k)=-d_{i}(-k)\;\;\;{\rm for}\;i=1,2,\;\;\;d_{3}(k)=d_{3}(-k)\;,
\end{eqnarray}
and hence only $d_{3}=M$ can be the mass term, and $\left\{d_{1},d_{2}\right\}$ must be of odd powers of $k$. In this class, the basis is the Bogoliubov quasiparticles 
\begin{eqnarray}
{\cal H}=\sum_{k}\left(c_{k}^{\dag}\;c_{-k}\right)
\left(
\begin{array}{cc}
M & d_{1}-id_{2} \\
d_{1}+id_{2} & -M
\end{array}
\right)\left(
\begin{array}{cc}
c_{k} \\
c_{-k}^{\dag}
\end{array}
\right)\;.
\label{1D_class_D_H_Bogoliubov}
\end{eqnarray}
To have a dispersion of the form $E_{\pm}(k)=\pm\sqrt{k^{2n}+M^{2}}$, we parametrize $d_{i}$ by
\begin{eqnarray}
d_{1}=a_{1}k^{n}\;,\;\;\;d_{2}=a_{2}k^{n}\;.
\end{eqnarray}
where $|a|=\sqrt{a_{1}^{2}+a_{2}^{2}}=1$ is rescaled to unity. We may further eliminate the $d_{1}$ component by a gauge transformation such that the Hamiltonian takes the form of the well-studied 1D Majorana chain\cite{Kitaev01}, and the correlation function can be introduced\cite{Molignini18}. We rewrite the Hamiltonian by 
\begin{eqnarray}
&&(d_{1}-id_{2})c_{k}^{\dag}c_{-k}^{\dag}=(a_{1}-ia_{2})k^{n}c_{k}^{\dag}c_{-k}^{\dag}
=|a|e^{i\varphi}k^{n}c_{k}^{\dag}c_{-k}^{\dag}
\nonumber \\
&&=-ik^{n}e^{i(\varphi+\pi)}c_{k}^{\dag}c_{-k}^{\dag}
=-i\tilde{d}_{2}\tilde{c}_{k}^{\dag}\tilde{c}_{-k}^{\dag}
\rightarrow -id_{2}c_{k}^{\dag}c_{-k}^{\dag}\;,
\end{eqnarray}
such that the Hamiltonian now reads 
\begin{eqnarray}
{\cal H}=\sum_{k}\left(c_{k}^{\dag}\;c_{-k}\right)
\left(
\begin{array}{cc}
M & -id_{2} \\
id_{2} & -M
\end{array}
\right)\left(
\begin{array}{cc}
c_{k} \\
c_{-k}^{\dag}
\end{array}
\right)\;,
\label{1D_class_D_H_after_gauge_trans}
\end{eqnarray}
with 
\begin{eqnarray}
d_{2}=k^{n}\;,\;\;\;n\in 2{\mathbb Z}+1\;,
\end{eqnarray}
which is now in a more convenient form.

For the case that the basis is fermionic, we introduce the Majorana basis by assuming that our continuous Hamiltonian is the low-energy effective theory of a lattice model, and introduce the Majorana fermions from the real space lattice electron creation and annihilation operators at site $i$\cite{Chiu16} 
\begin{eqnarray}
\lambda_{i}=c_{i}+c_{i}^{\dag}\;,\;\;\;\lambda_{i}^{\prime}=\frac{1}{i}(c_{i}-c_{i}^{\dag})\;,\;\;\;
\Lambda_{i}=\left(
\begin{array}{c}
\lambda_{i} \\
\lambda_{i}^{\prime}
\end{array}
\right)\;.
\end{eqnarray}
Then the Hamiltonian in Eq.~(\ref{1D_class_D_H_after_gauge_trans}) can be written as 
\begin{eqnarray}
&&{\cal H}=\frac{i}{4}\sum_{k}\Lambda^{T}(k)X(k)\Lambda(-k)
\nonumber \\
&&=\sum_{k}\Lambda^{T}(k)
\left(
\begin{array}{cc}
0 & d_{2/4}-iM/4 \\
d_{2}/4+iM/4 & 0 
\end{array}
\right)
\Lambda(-k)\;,
\nonumber \\
&&X(k)=-iM\sigma_{y}-id_{2}\sigma_{x}\;.
\label{1D_class_D_H_Majorana}
\end{eqnarray}
That is, $d_{2}/4$ is effectively the $\sigma_{x}$ component and $M/4$ the $\sigma_{y}$ component in this Majorana basis.

At HSPs $k_{0}=-k_{0}$, the PH symmetry $CH(k_{0})C^{-1}=-H(-k_{0})=-H(k_{0})$ implies that $d_{1}$ and $d_{2}$ in Eq.~(\ref{1D_class_D_H_Bogoliubov}) must vanish, as it does for our continuous model that sets $k_{0}=0$ at the origin. The $X(k_{0})$ in Eq.~(\ref{1D_class_D_H_Majorana}) is 
\begin{eqnarray}
X(k_{0}=0)=-iM\sigma_{y}=\left(
\begin{array}{cc}
 & -M \\
M & 
\end{array}
\right)\;,
\end{eqnarray}
which is antisymmetric. The ${\mathbb Z}_{2}$ topological invariant is calculated from the sign of the Pfaffian of $X(k_{0})$ at the two HSPs $k_{0}=0$ and $k_{0}=\pi$ in the original lattice model. In our continuous model
\begin{eqnarray}
{\rm Sgn}\left\{{\rm Pf}\left[X(k_{0}=0)\right]\right\}=-{\rm Sgn}(M)\;,
\end{eqnarray}
so there is a topological phase transition at $M_{c}=0$.

%{\cblue (1) Notice that I say in general the basis can be fermionic or bosonic, so I shouldn't just call it Majorana basis. For instance, for magnons it may represent $b+b^{\dag}\propto S_{x}$ or $b-b^{\dag}\propto S_{y}$. }

Alternatively, the topological invariant can be calculated from the Berry connection in the Majorana basis as in Sec.~\ref{sec:1D_class_BDI}. The eigenstates in the Majorana basis calculated from Eq.~(\ref{1D_class_D_H_Majorana}) is
\begin{eqnarray}
|u_{\pm}\rangle=\frac{1}{\sqrt{2}d}\left(
\begin{array}{c}
\pm d \\
d_{2/4}+iM/4
\end{array}
\right)\;,\;\;\;d=\frac{1}{4}\sqrt{d_{2}^{2}+M^{2}}\;.
\end{eqnarray}
The Berry connection of the filled band in this gauge is 
\begin{eqnarray}
A(k)=\langle u_{-}|i\partial_{k}|u_{-}\rangle
=\frac{M\partial_{k}d_{2}}{2\left(d_{2}^{2}+M^{2}\right)}\;.
\end{eqnarray}
After using $d_{2}=k^{n}$, we obtain the same form of $A(k)$ as in Eq.~(\ref{1D_class_BDI_Berry_connection}), and thus the same Zak phase ${\cal C}$ as in Eq.~(\ref{1D_class_BDI_Zak_phase}). We see that $\Delta{\cal C}=1$ regardless of the order of band crossing $n$. The correlation function follows from Eq.~(\ref{1D_class_BDI_correlation_fn}) which measures the overlap of Wannier function sandwiched by a position operator. Because the basis of this Wannier state is the Majorana fermions, the correlation function has been referred to as the Majorana-Wannier state correlation function, as previously calculated for the 1D Kitaev chain which is an explicit example for $n=1$ in this class\cite{Molignini18}. 
%Following the discussion we have shown in Sec.~\ref{sec:1D_class_BDI}, the correlation length has a critical exponent that is determined by the order of band crossing $\nu=1/n$. 
It should be reminded that although we formulate this secion in the fermionic language, our formalism also applies to bosonic models.

\subsection{1D class CII \label{sec:1D_class_CII}}

For 1D class CII with $T^{2}=-1$ and $C^{2}=1$, we use the $\Gamma$-matices\cite{Zhao14}
\begin{eqnarray}
\Gamma^{a}=\left\{\sigma_{x}\otimes\tau_{z},\sigma_{y}\otimes\tau_{z},
I\otimes\sigma_{x},I\otimes\tau_{y},\sigma_{z}\otimes\tau_{z}\right\}\;.
\end{eqnarray}
The TR and PH operators are $T=\sigma_{y}\otimes IK$ and $C=I\otimes\tau_{y}K$. 
The TR and PH symmetries require the mass term to be $d_{3}=M$, and the only allowed kinetic term is $d_{2}({\bf k})=-d_{2}(-{\bf k})$. Thus a general higher-order Dirac Hamiltonian in this class takes the form
\begin{eqnarray}
H=d_{2}\Gamma^{2}+d_{3}\Gamma^{3}=k^{n}\Gamma^{2}+M\Gamma^{3}\;,
\label{1D_class_CII_Hk}
\end{eqnarray}
with $n\in 2{\mathbb Z}+1$. The topological invariant is given by the winding number\cite{Zhao14} 
\begin{eqnarray}
{\cal C}=\frac{1}{2i}\int_{0}^{2\pi}\frac{dk}{2\pi}{\rm Tr}\left[\sigma_{y}\otimes\tau_{y}H^{-1}(k)\partial_{k}H(k)\right]\;,
\label{1D_class_CII_topo_invariant}
\end{eqnarray}
which in our continuous model of Eq.~(\ref{1D_class_CII_Hk}), with the replacement in the integral $\int_{0}^{2\pi}\rightarrow\int_{-\infty}^{\infty}$, yields ${\cal C}=-{\rm Sgn}(M)$, and hence the jump of topological invariant across the transition is $\Delta{\cal C}=2$ for any order of band crossing $n$.

The filled-band eigenstates read
\begin{eqnarray}
|u_{1-}\rangle=\frac{1}{\sqrt{2}d}\left(
\begin{array}{c}
-id_{2} \\
-d \\
0 \\
d_{3}
\end{array}
\right)\;,\;\;\;
|u_{2-}\rangle=\frac{1}{\sqrt{2}d}\left(
\begin{array}{c}
-d \\
id_{2} \\
d_{3} \\
0
\end{array}
\right)\;.
\end{eqnarray}
To construct the correlation function, we consider a modified Berry connection 
\begin{eqnarray}
\overline{A}^{ab}=\langle u_{a-}|\sigma_{y}\otimes\tau_{y}i\partial_{k}|u_{b-}\rangle\;,
\end{eqnarray}
such that the integrand in Eq.~(\ref{1D_class_CII_topo_invariant}) can be written as 
\begin{eqnarray}
{\rm Tr}\left[\sigma_{y}\otimes\tau_{y}H^{-1}(k)\partial_{k}H(k)\right]=4i{\rm Tr}\left[\overline{A}^{ab}\right]\;,
\end{eqnarray}
meaning that the topological invariant can as well be expressed as an integration of ${\rm Tr}\left[\overline{A}^{ab}\right]$. The Fourier transform of ${\rm Tr}\left[\overline{A}^{ab}\right]$ yields a correlation function
\begin{eqnarray}
\tilde{F}_{1D}^{\prime}(R)=\int_{0}^{2\pi}\frac{dk}{2\pi}e^{ikR}{\rm Tr}\left[\overline{A}^{ab}\right]={\rm Tr}\left[\langle 0|\sigma_{y}\otimes\tau_{y}{\hat r}|R\rangle\right]\;,
\end{eqnarray}
that measures the overlap of Wannier functions weighted by $\sigma_{y}\otimes\tau_{y}{\hat r}$.

%yields the matrix element of the TR operator
%\begin{eqnarray}
%m_{12}=\langle u_{1-}|T|u_{2-}\rangle=i\frac{d_{3}^{2}}{d^{2}}=i\frac{M^{2}}{k^{2n}+M^{2}}=-m_{21}\;.
%\end{eqnarray}
%This suggests we may use $-i\partial_{k}^{2}{\rm Pf}[m]=-i\partial_{k}^{2}m_{12}=F(k,M)$ as the curvature function, which yields a correlation function
%\begin{eqnarray}
%-i\tilde{F}_{TR}(R)=-i\int\frac{dk}{2\pi}e^{ikR}\partial_{k}^{2}{\rm Pf}\left[m(k)\right]=i\langle R1|R^{2}T|02\rangle\;.
%\end{eqnarray}
%The rest of the calculation follows that in Sec.~\ref{sec:3D_class_CII}.

%{\cblue (1) Redefine carefully, the correlation function has an extra factor of $-i$ compared to other ${\mathbb Z}_{2}$ correlation functions. }

\section{Conclusions \label{sec:conclusions}}

In summary, we investigate the quantum criticality of topological phase transitions for all the topologically nontrivial symmetry classes from 1D to 3D within a unified framework. Our approach generalizes the symmetry classification\cite{Schnyder08,Ryu10,Kitaev09,Chiu16} to higher-order Dirac models in all physically relevant cases, and reveals the following features due to the interplay between the topological invariant ${\cal C}$, the $\left\{T,C,S\right\}$ symmetries, order of band crossing $n$, and the $m$-fold rotational symmetry in 2D and 3D: (1) The $\left\{T,C,S\right\}$ symmetries constrain the order of band crossing to be even, odd, or integer in each dimension $\times$ symmetry class. (2) The even-oddness of the jump of topological invariant $\Delta{\cal C}$ at the critical point may be due to the even-oddness of the band crossing, or may also be because the formula for the topological invariant ${\cal C}$ only allows certain integer values. (3) The even-oddness of the band crossing generally gives a constraint for the rotational eigenvalues. As a result, it is possible that only certain rotational symmetries $m$ are compatible with a specific symmetry class, which may also depend on the system being fermionic or bosonic. (4) The topological invariant ${\cal C}$ (or a related one) takes the form of an integration over a curvature function. The Fourier transform of the curvature function always represents a correlation function that measures the overlap of two Wannier states that are a certain distance apart, or the overlap of multiple Wannier states in a compact form. (5) The critical exponent $\nu$ of the correlation length $\xi$, as well as that of the decay length of the edge state, are determined by the order of band crossing, but not necessarily $\Delta{\cal C}$ or $m$. (6) The critical exponent $\nu$ and that of the extremum of the curvature function $\gamma$ are not independent, but satisfy a scaling law owing to the conservation of the topological invariant. (7) The CRG approach based on the deformation of the curvature function can be used to judge topological phase transitions in all the dimensions and symmetry classes that have been investigated.

%(1) Note that the reason why each exponent is $2{\mathbb Z}$ or $2{\mathbb Z}+1$ is different in each class. For example in 3D class AIII the $\Delta C\in 2{\mathbb Z}+1$ is because the integration of topological invariant is only nonzero when  $\Delta C\in 2{\mathbb Z}+1$, whereas in 2D class C we have $\Delta C\in 2{\mathbb Z}$ because of the constraint from the nonspatial symmetry.

In the sense of clarifying how the $\left\{T,C,S,C_{m}\right\}$ symmetries influence the critical quantities $\left\{\Delta{\cal C},n,\nu,\gamma\right\}$, the present work brings the notion of universality class into the research of topological phase transitions, as summarized in Table \ref{tab:universality_class}. The identification of Wannier state correlation function further indicates that, despite lacking a local order parameter, one may still construct a correlation function at the wave function level to characterize the quantum criticality of the system, both in the topologically trivial and nontrivial phase. In addition, the CRG approach demonstrates that the concept of scaling is a process of deforming local curvature that leaves the global topology unchanged, in contrast to the coarse graining process for Landau order parameters.A significant byproduct of our analysis is the complete classification of Dirac models with TR, PH, chiral, and $m$-fold rotation symmetry. Importantly, this is also of relevance for the study of band crossing points in semimetals or superconductors that are protected by $m$-fold rotation. That is, symmetry-protected band crossings in semimetals or superconductors with $m$-fold rotation symmetry described by the considered Dirac models with momentum-dependent mass terms. We anticipate that this framework we follow can be generally applicable to study how other kinds of spatial symmetries, in conjunction with the symmetry classification according to $\left\{T,C,S\right\}$, influence the quantum criticality of topological phase transitions. Applications of this kind to a wide range of spatial symmetries, such as reflection or a specific point group symmetry, await further investigations.

%{\cblue (1) Need to mention which correlation function is measurable. }

\vspace{1cm}

\bibliographystyle{unsrt}
\bibliography{Literatur}

\end{document}